\documentclass[12pt]{article}
\usepackage{amsfonts}
\usepackage{amsthm}
\usepackage{amssymb}
\usepackage{amsmath}
\usepackage{graphicx,color,epstopdf}
\usepackage{hyperref}
\def\hDash{\bot\!\!\!\bot}

\newtheorem{theorem}{Theorem}[section]
\newtheorem{prop}{Proposition}[section]

\newtheorem{lemma}{Lemma}[section]
\newtheorem{remark}{Remark}[section]
\numberwithin{equation}{section}

\begin{document}
\title{Dimension reduction-based significance testing in nonparametric regression
\footnote{Lixing Zhu is a Chair professor of Department of Mathematics
at Hong Kong Baptist University, Hong Kong, China. He was supported by a grant from the
University Grants Council of Hong Kong, Hong Kong, China. }}
\author{Xuehu Zhu$^{1}$ and Lixing Zhu$^1$
\\$^1$Department of Mathematics, Hong Kong Baptist University, Hong Kong}
\date{}
\maketitle

\begin{abstract} \baselineskip=16pt
 A dimension reduction-based adaptive-to-model  test is proposed for  significance of a subset of covariates in the context of a nonparametric regression model. Unlike existing local smoothing significance tests, the new test behaves like a local smoothing test as if the number of covariates were just that under the null hypothesis and it can detect local alternatives distinct from the null at the rate that is only related to the number of covariates under the null hypothesis. Thus, the curse of dimensionality is largely alleviated when nonparametric estimation is inevitably required.  In the cases where there are many insignificant covariates, the improvement of the new test is very significant over existing local smoothing tests on the significance level maintenance and power enhancement.  Simulation studies and a real data analysis are conducted  to examine the finite sample performance of the proposed test.
\bigskip


\bigskip
{\it Running head.} Significance Testing
\end{abstract}
\newpage
\baselineskip=20pt

\newpage

\setcounter{equation}{0}
\section{Introduction}
Consider the  nonparametric regression model:
\begin{eqnarray}\label{(1.1)}
Y = m(Z) + \epsilon,
\end{eqnarray}
where $Y$ is a scale dependent variable with the covariates $Z=(X^{\top}, W^{\top})^{\top}$,  $X=(X_1,\cdots,X_{p_1})^{\top}\in \mathbb{R}^{p_1}$, $W =(W_1,\cdots,W_{p_2})^{\top} \in \mathbb{R}^{p_2} $ and $p_1+p_2=d$, the regression function $m(\cdot):\mathbb{R}^d \rightarrow \mathbb{R}$ is  unknown in its form and $\epsilon$ is the error term  with zero conditional expectation when $Z$ is given: $E(\epsilon|Z)=0.$  %As argued in Hausman (1978) that ''Specification tests form one of the most important areas for research in econometrics",
As well known, the success of any further statistical analysis hinges on the correction of working model. Note that in regression modeling, there are often part of the covariates to be redundant. A subset of the covariates $W$ is said to be insignificant for the response variable $Y$ given $X$ if
\begin{eqnarray}\label{(1.2)}
E(Y|X,W)=E(Y|X).
\end{eqnarray}
The equality (\ref{(1.2)}) means that  $W$ does not provide more information to predict $Y$.   $W$ should be removed from the regression model (\ref{(1.1)}), otherwise, such redundant variables cause statistical analysis more complicated and less accurate and efficient. This step, particularly in the first stage of regression analysis, is necessary. In the literature, relevant testing problem called significance testing
 has attracted much attention.
There exist several proposals that are based on prevalent local smoothing and global smoothing methodologies in the literature.   For the former, Lavergne and  Vuong (2000) extended the idea introduced by Fan and Li (1996), proposed a test based on a second conditional moment to check the significance of a subset of covariates.
Further, Li (1999) developed a nonparametric significance test that is based on  the idea of Fan and Li (1996)
for nonparametric and semiparametric time-series  models.  Racine et al. (2006) suggested  a test for the significance of categorical variables in fully nonparametric regression models.  Lavergne et al. (2014) devised a new kernel-based test that is based on a suitable equivalent Fourier transformation.
It is noted that these local smoothing-based test statistics converge to the respective limiting null distributions at the typical rate $O_p(n^{-1/2}h^{-d/4})$, where $d$ is the number of all the covariates and $h$ is the bandwidth in kernel estimation.  Further,  these tests can also only detect local alternatives distinct from the null hypothesis at the rate $O_p(n^{-1/2}h^{-d/4})$. When  $d=p_1+p_2$ is large, the convergence rate is very slow  because $h$ converges to zero at a certain rate.
This implies that these local smoothing methodologies severely suffer from the curse of dimensionality.  This problem is caused by using nonparametric estimation for the models under both the null and alternative hypothesis that assumes the significance of all the covariates. However, this  clearly has not yet fully used the model structure under the null hypothesis. An ideal way to construct a test as follows. The test only involves $p_1$ significant covariates under the null hypothesis such that we can have a much faster convergence rate to well maintain the significance level when the limiting null distribution is employed and enhance the power performance. For global smoothing tests,  Racine (1997)  advised a test based on  nonparametric estimation of partial derivative. Delgado and  Gonz\'alez Manteiga (2001) introduced a consistent test based on a stochastic process. Both the tests have a fast rate of order  $\sqrt{n}$.
Racine's (1997) test involves the nonparametric estimation of partial derivative  ${\partial E(Y|X, W)}/{\partial W}$ and thus, estimation inefficiency can  greatly deteriorate the performance of the test.
Delgado and  Gonz\'alez-Manteiga's (2001) test involves the multivariate nonparametric estimation of $E(Y|X)$ and all the covariates in the process. Thus, the data sparseness in high-dimensional space still causes negative impact for the power performance of the resulting test.  We can see from the following example that the power drops down very quickly as $d$ increases.
Finally, when the dimension $d$ of the covariates is large, the computational burden is also an issue  because the Monte Carlo approximations to their sampling null distributions are  computationally intensive.

%\begin{center}
%{\color{red} Dear Prof. Zhu,  the former part has been modified.}
%\end{center}

We now present the simulation results for the following illustrative example. The model is
%The above disadvantageous phenomena can be easily illustrated by the following simple linear model
\begin{eqnarray*}
Y = (X_1 + X_2)/\sqrt{2} +2W_1 + \epsilon,
\end{eqnarray*}
where $X=(X_1,X_2,\cdots, X_{p_1})^{\top}$ and  $W=(W_1,W_2,\cdots, W_{p_2})^{\top}$ respectively follow standard multivariate normal distributions with identity covariance matrices, the sample size is $n=200$ and the dimension of  $X$ is set to be $p_1=2,3, 4$ and the dimension $p_2$ of $W$ is chosen to be $1$ to $8$ in this numerical simulations.  For the presentation purpose in the figure, the power is set to be 1 when $p_2=0$. The hypothetical regression function is $(X_1 + X_2)/\sqrt{2}$. We use this example to examine how the powers of existing   tests drop down with increasing dimension $p_2$.
Here we choose Fan and Li (1996)'s test and Delgado and  Gonz\'alez Manteiga's test (2001) as the respective representatives of local and global smoothing tests to demonstrate this.

\begin{figure}
  \centering
  \includegraphics[width=7cm]{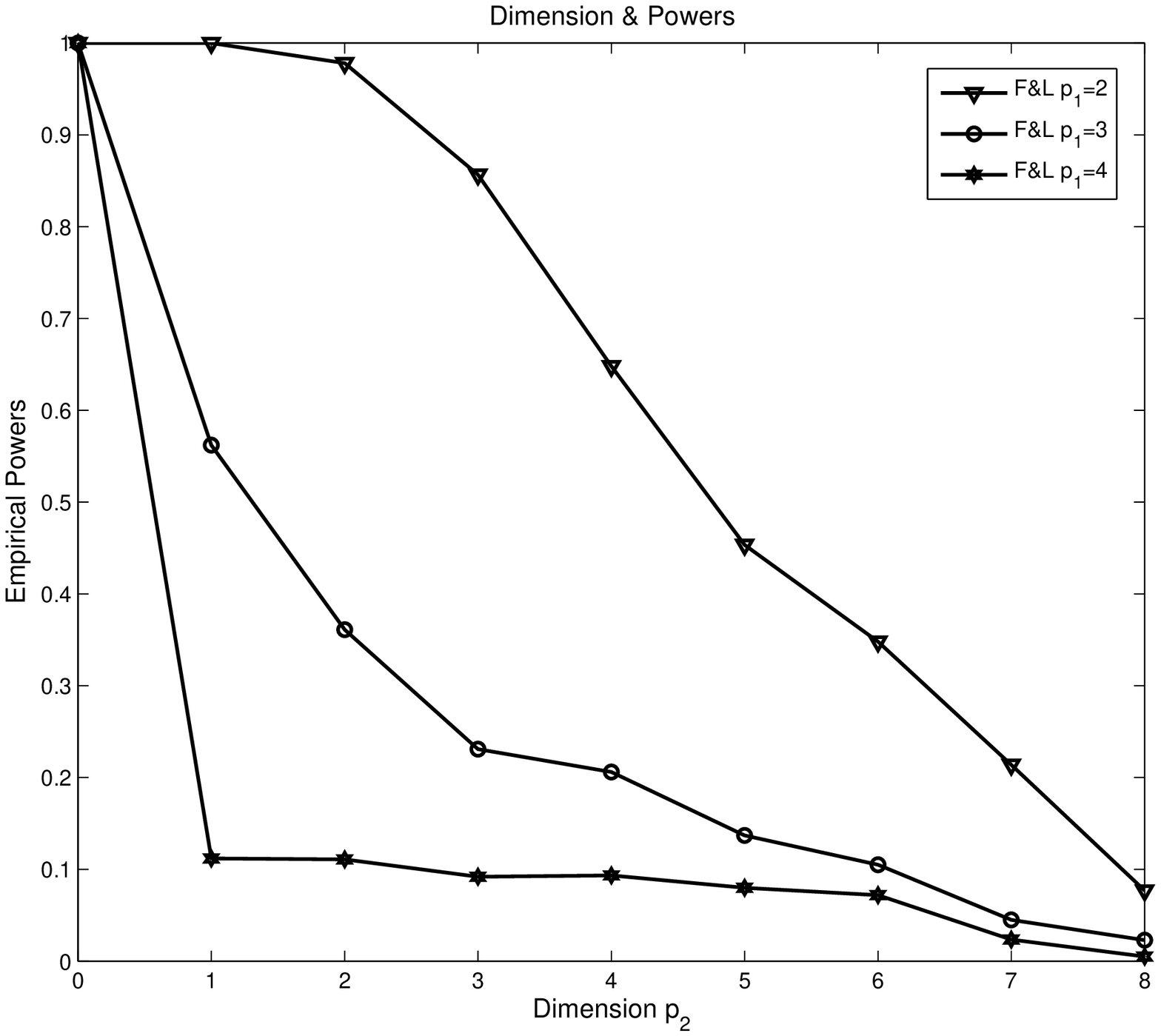}
    \includegraphics[width=7cm]{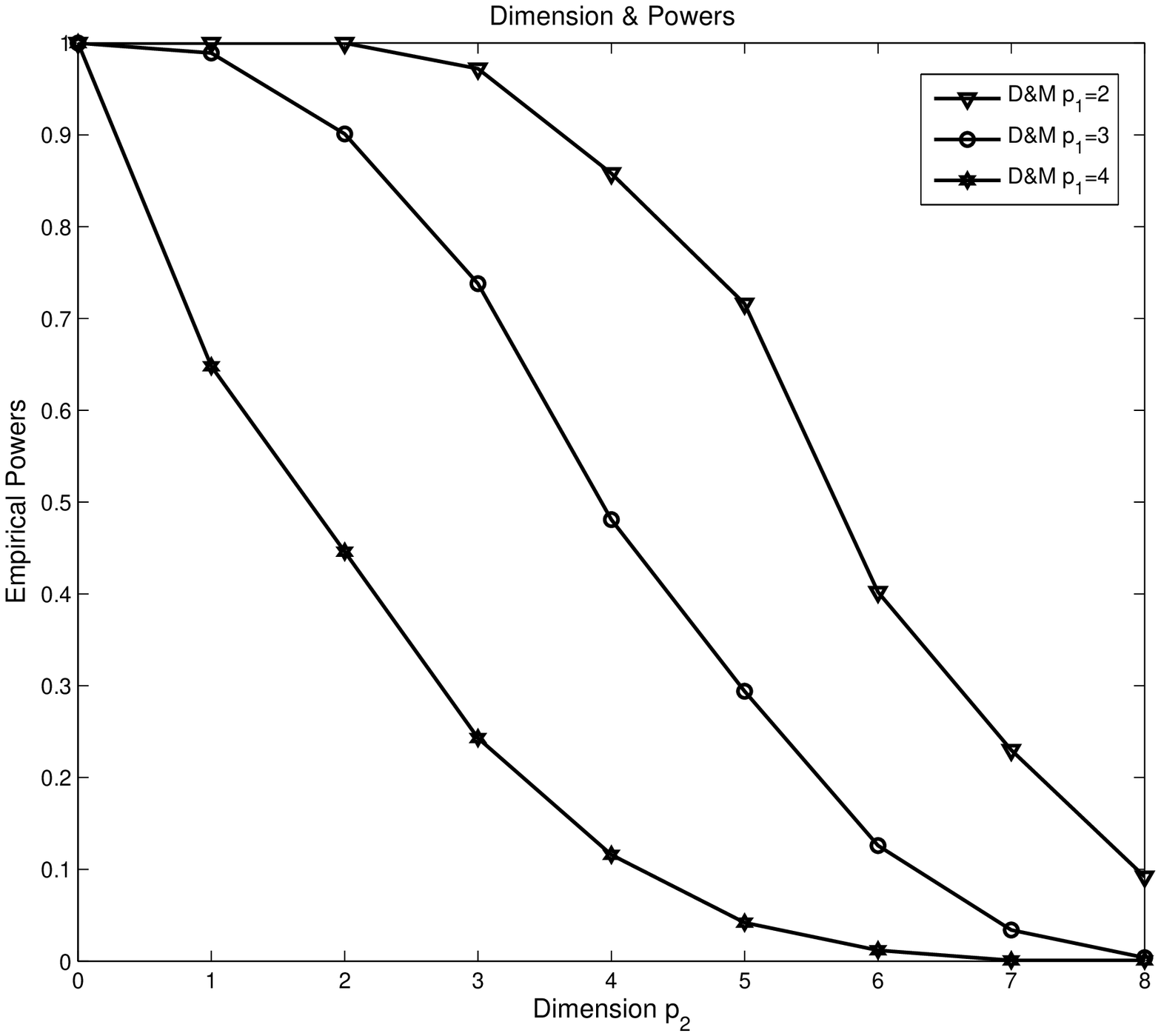} \\
  \caption{ The empirical power curves of Fan and Li's (1996) test and Delgado and Manteiga's (2001) test against the dimensions of $X$ and $W$ with sample size 200
  at the significance level $\alpha=0.05$.
           }\label{figure1}
\end{figure}

\begin{center}
%\begin{figure}
%  \centering
%  % Requires \usepackage{graphicx}
%  \includegraphics[width=7cm]{introductionfan05.eps}
%    \includegraphics[width=7cm]{introductionDM05.eps} \\
%  \caption{ The empirical power curves of Fan and Li's (1996) test and Delgado and Manteiga's (2001) test against the dimensions of $X$ and $W$ with sample size 200
%  at the significance level $\alpha=0.05$.
%           }\label{figure1}
%\end{figure}
Figures \ref{figure1}  about here
\end{center}
Figures \ref{figure1} depicts the curves of the empirical powers at the significance level $\alpha=0.05$. More  details about the bandwidth selection and other details for this simulation can be found in Section~4.  Obviously, the empirical powers of  these two tests rapidly decrease as $p_1$ and $p_2$ increase. This indicates that both the tests  severely  suffer from the curse of dimensionality. Relevant  discussions on the maintainance of the significance level will be discussed below.
To this end, how to overcome the aforementioned problems caused by dimensionality is of great
importance.

Guo et al. (2014) recently devised a dimension-reduction local
smoothing test which used to test generalized linear regression models. The
basic idea is to utilize the dimension reduction structure to adapt the true underlying regression models such that it behaves like a test with univariate covariate under the null hypothesis and adapts to the model structure to make the test omnibus under the alternative hypothesis. This approach greatly improves the performance of existing local smoothing tests on significance level maintainance and power enhancement. Under the null hypothesis, the test statistic converges to its limit at
a faster convergence rate $O(n^{-1/2}h^{-1/4})$ than the typical rate $O(n^{-1/2}h^{-d/4})$ of existing
local smoothing tests and can detect local alternative models distant from the null models at a much faster convergence rate $O(n^{-1/2}h^{-1/4})$.
Zhu et al. (2015b) followed the similar idea to develop a
dimension reduction global smoothing test for more general regression
models. Both of these adaptive methods can greatly overcome the curse of dimensionality.

It is of interest to apply the idea to construct a dimension reduction-based test in significance testing. However, to identify the projected covariates in the hypothetical and alternative models, they assumed that the projected covariates in the hypothetical models are contained in the  alternative models. In the present paper, we cannot impose this assumption and thus, the identification procedure  will be different in the model adaption step. The details will be seen from the following model and  the test statistic construction in the next section. As is known, the objective of significance testing focuses on choosing the significant covariates $X$ in the nonparametric regression setting. Let $g(x)=E(Y|X=x)$. Then the significance testing becomes the following hypothesis:
\begin{eqnarray}\label{(1.3)}
H_0:\  E(Y|X,W)=g(X) \quad \mbox{versus}\quad H_1: \  E(Y|X,W) \neq g(X)
\end{eqnarray}
Let $U=Y-g(X)$. Recall $Z=(X^{\top}, W^{\top})^{\top}.$ Then we have under the null hypothesis  $H_0$, $E(U|Z)=m(Z)-g(X)=0$, but under the alternative hypothesis $H_1$, $E(U|Z) \neq  0$. To facilitate a more general formulation of model structure we want to test, consider the following reformulation of the above model.
Note that $g(\cdot)$ is an unknown function, the nonparametric regression model $Y =g(X) + U$ can be reformulated as:
\begin{eqnarray*}
Y =g(X) + U=g(B_1B^{\top}_1X) + U \equiv:  \tilde{g}(B^{\top}_1X) + U ,
\end{eqnarray*}
where $B_1$ is an orthonormal $p_1\times p_1$ matrix. This means that the above regression model  can be viewed as a special multi-index regression model with $p_1$ indices corresponding to the covariates $X$.
Thus, in the present paper, we consider a more general null hypothesis as:
\begin{eqnarray}\label{(1.4)}
H_0: E(Y|Z)=g(B^{\top}_1X)=E(Y|X)\quad \rm a.s,
 \end{eqnarray}
where $B_1$ is a $p_1 \times q_1$ orthonormal matrix in the sense that $B^{\top}_1B_1=I_{q_1\times q_1}$.   $q_1$ with $1\leq q_1 \leq p_1$ is assumed to be given.
The hypothetical regression model covers many popularly used models in the literature, including the single-index models, the multi-index models and the partially linear single-index models. When the above regression model is a single-index model or partially linear single index model, the corresponding number $q_1$ of the indices
becomes one or two, respectively. Particularly, when $q_1= p_1$, the hypothetical model becomes the classical model of  (\ref{(1.3)}).
The alternative hypothesis now is
\begin{eqnarray*}
 E(Y|Z)=m(B^{\top}_1X, W) \neq g(B^{\top}_1X)=E(Y|X).
 \end{eqnarray*}
The term $m(B^{\top}_1X, W)$ can be reformulated as:
\begin{eqnarray*}
 E(Y|Z)=m(B^{\top}_1X, W) =m(B^{\top}Z),
 \end{eqnarray*}
where
\begin{eqnarray*}
 B= \left( \begin{array}{ll}
B_1 & 0_{p_1\times p_2}\\
0_{p_2\times q_1} & I_{p_1\times p_1}\\
\end{array} \right).
 \end{eqnarray*}
%Under the null hypothesis, the regression (\ref{(1.1)}) be rewritten as:
%\begin{eqnarray*}
% E(Y|Z)=g(B^{\top}_1X)= g(B^{\top}Z)=E(Y|X)\quad \rm a.s,
% \end{eqnarray*}
%where $B=(B^{\top}_1, O^{\top}_{p_2 \times q_1})^{\top}$.
From this reformulation, we similarly consider a more general alternative hypothesis as:
\begin{eqnarray}\label{(1.5)}
H_1: E(Y|Z)=m(B^{\top}Z) \neq g(B^{\top}_1X)=E(Y|X),
 \end{eqnarray}
where $B$ is   a $d\times q$  matrix with  $d \ge q>q_1$ being an unknown number and $E(\varepsilon|Z)=E(\varepsilon|B^{\top}Z)=0$.
 %where $B$ is   a $d\times q$  matrix $q$  with  $q=q_1+q_2\le d$ where $q_2$ is an unknown value and $E(\varepsilon|Z)=E(\varepsilon|B^{\top}Z)=0$.
For identifiability, assume $B^{\top}B=I_{q\times q}$.
Therefore, in this paper, we test the null hypothesis~(\ref{(1.4)}) against the alternative hypothesis~(\ref{(1.5)}).

To this end,  we develop a test that  utilizes the advantage with less dimension under $H_0$ and automatically adapt to $H_1$ such that the test is omnibus. We will call it a Dimension REduction Adaptive-to-Model test (DREAM). Compared with existing local smoothing tests,  we will show that DREAM converges to their limit at the rate of order $O(n^{-1/2}h^{-q_1/4})$ and can detect the local alternatives distinct from the null also at this rate, rather than at the rate of order $O(n^{-1/2}h^{-d/4})$.  Further, the critical values of the new test can be determined by its limiting null distribution. It is worthwhile to mention that almost all existing local smoothing tests require the assistance of Monte Carlo approximation to determine critical values otherwise, the significance level is difficult to maintain. However, the dimension reduction structure of the test alleviates this difficulty. More details will be discussed in the next sections.

This paper is organized as follows. In Section 2, the test statistic construction is described. Because  dimension reduction technique plays a very important role, we first briefly review a promising method: the discretization-expectation estimation. To make the test have the model-adaption property, a ridge-type eigenvalue ratio estimate (RERE) for the dimension $q$ of $B$ is recommended, which can consistently estimate $q_1$ and $q$ under the null and  alternative hypotheses accordingly. The asymptotic properties of the test statistic are presented in Section 3. Further, the test statistic tends to infinity at the certain rate under the global alternative hypothesis in Section 3.
In Section 4, we examine the finite sample performance of our test and also apply it to a real data analysis for illustration.
All the technical conditions and the proofs of the theoretical results are postponed to the Appendix.

\section{Dimension reduction-based adaptive-to-model test}

\subsection{Basic test statistic construction}
%Let $\tilde{B} = (B^{\top}_1, O^{\top}_{p_2 \times q_1})^{\top}$. The null and alternative hypothesis can be rewritten as:
%\begin{eqnarray}\label{(2.1)}
%H_0: E(Y|Z)=g(B^{\top}_1X),\ a.s, \quad \mbox{versus}\quad H_1:  E(Y|Z)=m(B^{\top}Z)\neq g(\tilde B^{\top}Z).
%\end{eqnarray}
%In other words,
%under the null hypothesis, $B$ is reduced to $\tilde{B}$ that has $q_1$ columns.

%It is  worth noticing that in the above formulations, $B$ and $B_1$ are usually not identifiable. Actually, for any $q\times q$ orthogonal matrix $C$, under $H_1$, $E(Y|Z)=E(Y|B^{\top}Z)=E(Y|CB^{\top}Z)$ and for any $q_1\times q_1$ orthogonal matrix $ C_1$, under $H_0$, $E(Y|Z)=E(Y|B_1^{\top}X)=E(Y| C_1 B_1^{\top}X)$. Hence, what we can identify is just $BC^{\top}$  for a matrix $C$ under $H_1$ and $B_1 C_1^{\top}$ for a matrix $ C_1$ under $H_0$. However, it is enough to have such a weaker identification because  $C_1 B_1^{\top}Z$ does not get $W$ involved. In the following subsection, we will briefly introduce a method to identify $BC^{\top}$  for a matrix $C$ and $ B_1\tilde C^{\top}$ for a matrix $C_1$.
%Without notational confusion, we use $B$ to  write $BC^{\top}$, or $B_1C_1^{\top}$  throughout the rest of the present paper.

It is  worth noticing that in the above formulations, $B$ and $B_1$ are usually not identifiable. Actually, for any $q\times q$ orthogonal matrix $C$, $E(Y|Z)=E(Y|B^{\top}Z)=E(Y|CB^{\top}Z)$ and for any $q_1\times q_1$ orthogonal matrix $ C_1$, $E(Y|X)=E(Y|B_1^{\top}X)=E(Y| C_1 B_1^{\top}X)$. Hence, what we can identify is  $BC^{\top}$  for a matrix $C$  and $B_1 C_1^{\top}$ for a matrix $ C_1$.
Under $H_0$, since $E(Y|Z)=E(Y|B^{\top}Z)=E(Y|B_1^{\top}X)$, $B$ is automatically reduced to a $d\times q_1$ matrix $\tilde B=(B_1^{\top}, O_{q_1\times p_2}^{\top})^{\top}$ and $q=q_1$.
Therefore, it is enough to have such a weaker identification because  $C B^{\top}Z=C (B_1^{\top}, O_{q_1\times p_2}^{\top})^{\top}Z=CB_1^{\top}X$ still does not involve $W$. In the following subsection, we will briefly introduce a method to identify $BC^{\top}$  for a matrix $C$ and $ B_1 C^{\top}_1$ for a matrix $C_1$.
Without notational confusion, we use $B$ and $B_1$ to write $BC^{\top}$ and $B_1C_1^{\top}$, respectively,   throughout the rest of the present paper.

Under the null hypothesis $H_0$, we have
$$E(U|Z)=0 \Longrightarrow E(U| Z)
=E(\epsilon| B^{\top}Z)= 0.$$
Then
\begin{eqnarray}\label{(2.2)}
&&E(U E(U|B^{\top}Z)W(B^{\top}Z))=E(E^2(\epsilon|B^{\top}Z)W(B^{\top}Z))=0,
\end{eqnarray}
where $W(B^{\top}Z)$ is some positive weight function that will be  discussed below.
Under the alternative hypothesis $H_1$, since
$$E(U|B^{\top}Z)=m(B^{\top}Z)- g(B^{\top}_1X)\neq 0,$$
 we have
\begin{eqnarray}\label{(2.3)}
E(U E(U|B^{\top}Z)W(B^{\top}Z))=E(E^2(U|B^{\top}Z)W(B^{\top}Z)) >0.
\end{eqnarray}
The above argument implies that the empirical version of the left hand side in
(\ref{(2.2)}) can be viewed as a base to construct a test statistic. Further, the null hypothesis $H_0$ is rejected for large values of the test statistic. This motivates a naive construction as any one in the literature. However, we also note that under the null hypothesis,
\begin{eqnarray}\label{(2.2n)}
&&E(U E(U|B^{\top}Z)W(B^{\top}Z))=E(E^2(\epsilon|B_1^{\top}X)W(B_1^{\top}X))=0.
\end{eqnarray}
This means that under the null hypothesis, the dimension is reduced from $q$ to $q_1$.

The key is  how to construct a test statistic that fully uses this piece of information and can automatically adapt the model structure under the alternative hypothesis such that the test is still omnibus. We will present our idea in the following construction.

When a sample $\{(z_1, y_1), \cdots, (z_n, y_n)\}$ is available, the residual term $u_i$ is estimated as $\hat{u}_i = y_i-\hat{g}(\hat{B}^{\top}_1x_i)$, where $\hat{g}(\hat{B}^{\top}_1x_i)$ is a kernel estimate of $g(B^{\top}_1x_i)$ as following form:
\begin{eqnarray*}
\hat{g}(\hat{B}^{\top}_1x_i) = \frac{\frac{1}{n-1}\sum_{j\neq j}^n
Q_{h_1}(\hat{B}^{\top}_1x_j-\hat{B}^{\top}_1x_i)y_j}{\frac{1}{n-1}\sum_{j\neq j}^n
Q_{h_1}(\hat{B}^{\top}_1x_j-\hat{B}^{\top}_1x_i)},
\end{eqnarray*}
and $\hat{p}_{\hat{B}_1i}$ is a kernel estimate of the density function $p_{B_1}(\cdot)$ of $B^{\top}_1 x_i$ given by
\begin{eqnarray*}
\hat{p}_{\hat{B}_1i}=\frac{1}{n-1}\sum_{j=1}^n
Q_{h_1}(\hat{B}^{\top}_1x_j-\hat{B}^{\top}_1x_i),
\end{eqnarray*}
and where $Q_{h_1}=Q(\cdot/h_1)/h^{q_1}_1$ with $Q(\cdot)$ being a $q_1$-dimensional
product kernel function from the univariate kernel $\tilde{Q}(\cdot)$, $h_1$ being a bandwidth and $\hat{B}_1$ being an estimate of $B_1$.
Then we obtain the following kernel estimate $E(U|B^{\top}Z)$ as:
\begin{eqnarray*}
&&\hat E(\hat{u}_i|{\hat B}^\top
z_i)=\frac{\frac{1}{n-1}\sum_{j\neq i}^n
 K_{\hat{q}h}(\hat{B}^{\top}z_j-\hat{B}^{\top}z_i)\hat{u}_j}{\frac{1}{n-1}\sum_{j\neq i}^n
 K_{\hat{q}h}(\hat{B}^{\top}z_j-\hat{B}^{\top}z_i)}.
\end{eqnarray*}
In this formula, $\hat{B}$ is an estimate of $B$ with an estimated structural dimension $\hat{q}$ in a certain sense that will be specified in the next subsection, where $K_{\hat{q}h}=K(\cdot/h)/h^{\hat{q}}$ with $K(\cdot)$ being a $\hat{q}$-dimensional kernel function and $h$ being a bandwidth.  If we choose the weight $W(\cdot)$ to be the density function $p_{B}(\cdot)$ of $B^{\top}Z$, for any ${\hat B}^{\top}z_i$, we can estimate $p_{B}(\cdot)$ in the following form:
\begin{eqnarray*}
\hat{p}_{\hat{B}i}=\frac{1}{n-1}\sum_{j\neq i}^n
 K_{\hat{q}h}(\hat{B}^{\top}z_j-\hat{B}^{\top}z_i).
\end{eqnarray*}
Therefor, a non-standardized test statistic is defined by
\begin{eqnarray}\label{(2.4)}
V_{n}&=&\frac{1}{n(n-1)}\sum_{i=1}^n\sum_{j\neq i}^n \hat
u_i\hat u_j K_{\hat{q}h}(\hat{B}^{\top}z_j-\hat{B}^{\top}z_i).
\end{eqnarray}

\begin{remark}
Note that  the test statistic developed by Fan and Li (1996) is:
\begin{eqnarray}\label{(2.5)}
\tilde{V}_{n}=\frac{1}{n(n-1)}\sum_{i=1}^n\sum_{j\neq i}^n \hat
u_i\hat{f}_{1i}\hat u_j\hat{f}_{1j}\tilde K_h(z_i-z_j),
\end{eqnarray}
where $\tilde{K}_h(\cdot)=K(\cdot/h)/h^p$ with $K(\cdot)$ being a $(p_1+p_2)$-dimensional kernel function and $\hat{f}_{1}$ is an estimate of the density function $f_{1}(\cdot)$ of $X$ no matter whether the underlying model is under the null or alternative hypothesis. Compared the formula (\ref{(2.4)}) with (\ref{(2.5)}), the difference is that our test  uses ${\hat B}^\top Z$ in lieu of $Z$ and applies $K_h(\cdot )$ in $V_n$ instead of $\tilde K_h(\cdot)$. At first glance, this difference seems not fundamental. {\it However, we just want to use an estimate of $B$ to make the test  adapt to the underlying models under the null and alternative hypothesis.} Thus, how to construct an estimate of $B$ plays a crucial role for model-adaption. The requirement is that an estimate of $B$ must have the following property: under $H_0$, it estimates $\tilde B=(B_1^{\top}, O_{q_1\times p_2}^{\top})^{\top}$ and under $H_1$ it automatically estimates $B$.
To achieve this goal, we also need an estimate $\hat q$ that can be consistent to $q_1$ under $H_0$ and to  $q$ under $H_1$. We will see that a standardizing version $nh^{q_1/2}V_{n}$ can be used  such that it has a finite limit under $H_0$ and diverges to infinity  much faster than  $nh^{(p_1+p_2)/2}\tilde V_n$ in Fan and Li's test statistic. The results are reported in Section~3.
\end{remark}

\subsection{A brief review on discretization-expectation estimation}
As we commented above, we need to identify $BC^{\top}$ and $B_1 C_1^{\top}$.  To this end, a method  is discussed in this subsection. The method is to identify the spaces spanned by $B$ and $B_1$ automatically under the null and alternative hypotheses. In other words, the method is to identify the  basis vectors in the respective subspaces.
This is an estimation problem  for the central mean subspaces in sufficient dimension reduction ( e.g. Cook 1998). The respective central mean subspaces are respectively denoted as
$\emph{S}_{E(Y|X)}$ and $\emph{S}_{E(Y|Z)}$. Also, $q_1= dim(\emph{S}_{E(Y|X)})$ and $q=dim(\emph{S}_{E(Y|Z)})$ are respectively called the structural dimensions of  $\emph{S}_{E(Y|X)}$ and $\emph{S}_{E(Y|Z)}$.
Here we assume that $q_1$ is known, but $q$ is unknown.

There exist several dimension reduction proposals available in the literature. For example, Li (1991) proposed  sliced inverse regression (SIR), Cook and Weisberg (1991) advised sliced average variance estimation (SAVE), Xia et al. (2002) discussed minimum average variance estimation (MAVE), Li and Wang (2007) presented directional regression (DR).
Cook and Forzani (2009) developed likelihood acquired directions (LAD), Zhu et al. (2010a) suggested discretization-expectation estimation (DEE) and Zhu et al. (2010b) provided average partial mean estimation (APME).
 In this paper,  we adopt DEE because it is computationally inexpensive without any tuning parameter selection that is required by SIR, SAVE or DR, and can be easily used to construct a criterion for determining $q$.

From Zhu et al. (2010a),  to identify and estimate $B$, the   DEE estimation procedure can be summarized as the following steps.
\begin{enumerate}
\item Define the set of binary variables $\Upsilon(t)= I\{Y \leq t\}$ by discretizing the response variable $Y$, where the indicator function $I\{Y \leq t\}=1$ if $Y \leq t$ and 0, otherwise.
\item Let $\emph{S}_{\Upsilon(t)|Z}$ denote the central subspace of $\Upsilon(t)|Z$, and $M(t)$ be an $ d \times  d$ positive semi-definite matrix satisfied $\rm{Span}\{M(t)\} =\emph{S}_{\Upsilon(t)|Z}$.
\item Let $\tilde{Y}$ denote an independent copy of $Y$. Taking the expectation over the random variable $\tilde{Y}$,  the target matrix becomes $M=E\{M(\tilde Y)\}$. $B$ consists of the eigenvectors associated with the nonzero eigenvalues of $M$.
\item Get an estimation of the target matrix $M$ as:
\begin{equation*}
M_{n}=\frac{1}{n}\sum^{n}_{i=1}M_n({y}_i),
\end{equation*}
where $M_n({y}_i)$ is the estimating matrix of $M({y}_i)$ by some certain sufficient dimension reduction method such that SIR. Then the estimate $\hat{B}$ of $B$ consists of the eigenvectors associated with the largest $q$ eigenvalues of $M_{n}$. Virtually, $\hat{B}$ is root-$n$ consistency of the matrix $B$ when $q$ is given.
\end{enumerate}
In this paper, the target matrices $M(t)$ and $M$ are based on  sliced inverse regression (SIR). More details can be referred to Li (1991) or Zhu et al. (2010a). Similarly, we can also utilize the DEE procedure to get an estimate $\hat{B}_1$ of the matrix $B_1$.
%Next, we consider estimating $q$.

The following proposition states the consistency of the estimated matrix $\hat{B}$ under $H_0$.
\begin{prop}\label{Bnnull}
Under $H_0$ and Conditions A1 and A2 in Appendix, the DEE-based estimate $\hat{B}$  is consistent to $\tilde B= (B_1^{\top}C_1, O_{q_1\times p_2}^{\top})^{\top}$ for  some $q_1\times q_1$ orthogonal matrix $C_1$.
\end{prop}

Proposition \ref{Bnnull} indicates that under $H_0$, in a probability sense the test statistic only uses those variables that are significant. The curse of dimensionality can then be largely alleviated when nonparametric estimation is inevitably required.

However, as $q$ is unknown, to accommodate the alternative hypothesis we should estimate it consistently.  Thus, we need to estimate $q$ in general even  under $H_0$ with a given  $q=q_1$ such that  we can then define a final estimate of $B$. The following subsection provides an estimate and its consistency.

%{\color{red}However, the above algorithm cannot automatically adapt to the underlying model such that $B$ and $B_1$ can be automatically identified. This is because under the hypothetical model, the covariates are $X$ rather than $Z$ and if we  implement the above algorithm, the target matrix must be $p_1\times p_1$ instead $d\times d$ that is with the covariates $Z$. The problem is that we do not know whether the underlying model is under $H_0$ or $H_1$, and whether the corresponding structural dimension is $q$ or $q_1$. To attack this problem, we use the following reformulation of the hypothetical model. Let $\tilde B=(B_1^{\top}, O_{q_1\times p_2}^{\top})^{\top}$ where $O_{q_1\times p_2}^{\top}$ is a zero matrix. We now rewrite the hypothetical model as $E(Y|Z)=m(B_1^{\top}X)=g(\tilde B^{\top}Z).$ In this reformulation, $\tilde B$ is a $d\times q_1$ matrix. Thus, we can use $(Z, Y)$ to construct the DEE-based target matrix that is $d\times d$ under both the null and alternative hypothesis. the key structure is that under the null, the structural dimension is $q_1$ while under the alternative, it becomes $q$. When an estimate $\hat q$ is consistent to $q_1$, we then understand that  a matrix $\hat B_1$ can be defined by using the first $\hat q$ rows of $\hat B$ and the resulting $\hat {\tilde B}=(\hat B_1^{\top}, O_{\hat q_1\times p_2}^{\top})^{\top}$. The following subsection presents a method to estimate both $q_1$ and $q$ and then to estimate $B_1$ and $B$.}

\subsection{Structural dimension estimation}
We define a criterion to estimate $q$ and $q_1$ in an automatic manner. Although the BIC-type criterion in Zhu et al. (2010a) that was motivated from  Zhu et al. (2006) is a candidate, choosing  an appropriate penalty is a difficult issue.

In this paper, we recommend a ridge-type eigenvalue ratio estimate (RERE). Based on our experience in practice, it is not very sensitive to the ridge choice. Let $\hat{\lambda}_{d} \leq \hat{\lambda}_{d-1}\leq \cdots \leq \hat{\lambda}_{1}$ be the eigenvalues of the estimating matrix $M_n$. Define
\begin{eqnarray}\label{(lambda*)}
\lambda^*_{i}=\frac{\hat{\lambda}_{i}-\frac{1}{\sqrt{n}}}{\hat{\lambda}_{i}-\frac{1}{\sqrt{n}}+1}, \ \mbox{for} \ 1\leq i \leq d.
\end{eqnarray}
Theoretically, we can use the original  We use $\lambda_{i}$ to define the following ratio-based criterion. However, we found that a couple of largest eigenvalues tend to be much larger than the other non-zero eigenvalues. Then some  ratios of estimated non-zero eigenvalues could be smaller than the minimizer, and then the structure dimension $q$ is often underestimated. Thus, we  `standardize' the eigenvalues $\lambda^*_i$ to define a criterion and estimator:
\begin{eqnarray}\label{(2.7)}
\hat{q}=\arg\min_{1\leq j \leq d}\left\{j: \frac{(\lambda^*_{j+1})^2+c_n}{(\lambda^*_j)^2+c_n}\right\}.
\end{eqnarray}
This method is motivated by Xia et al. (2015) and Zhu et al. (2015a). This algorithm is easily implemented.

The following proposition states the estimation consistency.
\begin{prop}\label{prop1}
In addition to  Conditions A1, A2 and A3 in Appendix, assume $c\times \log{n}/{n} \leq c_n \rightarrow 0$ with some fixed $c>0$. Then the estimate $\hat{q}$  in (\ref{(2.7)}) is consistent to $q_1$ under $H_0$ and to $q$ under $H_1$.
\end{prop}

From  Proposition \ref{prop1}, the choice of $c_n$ can be in a relatively wide range to guarantee the estimation consistency under the null and global alternative hypothesis.

 Altogether, the final estimate $\hat{B}_{\hat q}$ of $B$ can have the model adaptive property in the sense that under $H_0$, it is consistent to $\tilde B= (B_1^{\top}C_1, O_{q_1\times p_2}^{\top})^{\top}$ for a $q_1\times q_1$ orthogonal matrix $C_1$ and under $H_1$, to $B$.

%Further, we will prove that $\hat{q}$ converges to $q_1$, rather than $q$, under the sequence of local alternative models converging to the hypothetical model. In other words, under the local alternatives that we will consider below, the estimate $\hat q$ is not consistent to the true dimension $q$. However, this inconsistency has a positive impact for detecting local alternative models. We will discuss this in detail.
%{\color{red} Combining the estimation of $B$ and the relevant structural dimension $q$, we can see that the estimate $\hat B_{\hat q}$ can have the model adaptive property in the sense that under $H_0$, it is consistent to $B_1$ and under $H_1$, to $B$.}
\section{Asymptotic properties}
\subsection{Limiting null distribution}
First, define some notations. Let
\begin{eqnarray}\label{(3.1)}
s^2 = 2\int{}K^2(u)duE\{[Var(\varepsilon^2|B^{\top}Z)]^2p(B^{\top}Z)\},
\end{eqnarray}
and
\begin{eqnarray}\label{(3.2)}
\hat{s}^2 = \frac{2}{n(n-1)}\sum_{i=1}^n\sum_{j \neq i}^nK^2_h\left(\hat{B}^{\top}z_i-\hat{B}^{\top}z_j\right)\hat{u}^2_i\hat{u}^2_j.
\end{eqnarray}

\begin{theorem}\label{the1}Under $H_0$ and the regularity conditions in Appendix, we have
\begin{eqnarray*}
n h^{q_1/2}V_n \Rightarrow N(0, {s^2}).
\end{eqnarray*}
Further, ${s^2}$ can be  consistently estimated  by
$\hat{s}^2$.
\end{theorem}
Therefor, according to Theorem \ref{the1}, we  can get the standardized test statistic as:
\begin{eqnarray*}
T_{n} =  nh^{q_1/2}V_{n}/\hat{s}.
\end{eqnarray*}
Further, applying the Slusky theorem yields that under $H_0$ $T_n$ is asymptotically normal:
\begin{eqnarray*}
T_{n} \Rightarrow N(0, 1).
\end{eqnarray*}

\subsection{Power study}
We now study the power performance of the test statistic $T_{n}$. Consider the following sequence of local alternative hypotheses as:
\begin{eqnarray}\label{(3.3)}
H_{1n}:\ E(Y|Z)=g(B^{\top}_1X)+ C_n G(B^{\top}Z).
\end{eqnarray}
Fixed $C_n$ corresponds to the global alternative model and when $C_n$ goes to zero, the sequences are local alternative hypotheses.

To obtain the main results about the power performance under $H_{1n}$ of (\ref{(3.3)}), we first present the asymptotic behavior of the estimate $\hat{q}$ when $C_n \to 0$.  It is noted that although the structural dimension of the models is $q$, the estimate $\hat{q}$ converges to $q_1$ rather than  $q$. This is caused by the convergence of the local alternative models to the hypothetical model as $n \rightarrow \infty $, the following lemma states the result.

\begin{lemma}\label{lemma1} Under $H_{1n}$ of (\ref{(3.3)}) with $C_n = n^{-\frac{1}{2}}h^{-\frac{q_1}{4}}$ and the regularity conditions in Proposition~\ref{prop1} except that $c\times C^2_n\log{n} \leq c_n \rightarrow 0$ for some fixed $c>0$, as $ n\rightarrow 0 $, $\hat{q}$ determined by (\ref{(2.7)}) converges to $q_1$ with a probability going to 1.
\end{lemma}
It is interesting that this inconsistency even has a positive impact for detecting local alternative models.
The following theorem describes how sensitive the test is to the local alternative models.
\begin{theorem}\label{the2} Under the  regularity  conditions in Appendix, we have the following results.
\begin{itemize}
\item [(i)] Under  $H_{1n}$ with a fixed $C_n >0$
\begin{eqnarray*}
T_{n}/(n h^{q_1/2}) \Rightarrow {Constant} >0.
\end{eqnarray*}
\item [(ii)] Under  $H_{1n}$ with $C_n = n^{-\frac{1}{2}}h^{-\frac{q_1}{4}}$
\begin{eqnarray*}
n h^{q_1/2}V_n \Rightarrow N(u, {s^2}),
\end{eqnarray*}
where $u=E\big([E\{G(B^{\top}Z)|B_1^{\top}X\}]^2p_{B_1}(B_1^{\top}X)\big)$ and $s^2$ is given by (\ref{(3.1)}). Further,
${s^2}$ can be consistently estimated by  $\hat{s}^2$.
\end{itemize}
\end{theorem}
\begin{remark}
The results in this theorem confirm our claim in the first section. The convergence rate of the test statistic is $n h^{q_1/2}$ and the test can detect  local alternative models converging to the hypothetical model also at the rate of order $C_n = n^{-\frac{1}{2}}h^{-\frac{q_1}{4}}$. Fan and Li's (1996) test, which is also the case for existing local smoothing tests, can have the respective rates where $q_1$ is replaced by $d$, which causes a much slower rate.
\end{remark}

\section{Numerical Studies}
\subsection{Simulations}
In this subsection, we conduct the simulations to investigate the finite sample performance of our proposed test.
The empirical sizes and powers are computed via  2000 replications of the experiments at the significance level $\alpha = 0.05$.   Write the DREAM test as $T^{DEE}_{n}$. For comparison, we use Fan and Li's (1996) test and Delgado and Gonz\'alez Manteiga's (2001) test as the representatives of existing local and global tests. Write them as $T^{FL}_n$ and $T^{DM}_n$.

Delgado and Gonz\'alez Manteiga's (2001) test is defined as:
\begin{eqnarray*}
\tilde{V}_{n}=\frac{1}{n(n-1)}\sum_{i=1}^n[\sum_{j\neq i}^n \hat{u}_j\hat{f}_{1j}I(x_j<x_i)I(w_j<w_i)]^2.
\end{eqnarray*}
The critical values are determined by the wild bootstrap.  The bootstrap observations are from :
$y^*_i= \hat{g}(x_i) +\hat{u}_i\times V_i$,
where  $\{V_i\}_{i=1}^n$ is a sequence of i.i.d. random variables from the two-point distribution as:
\begin{eqnarray*}
P(V_i=\frac{1-\sqrt{5}}{2})=\frac{1+\sqrt{5}}{2\sqrt{5}},\qquad\,\, P(V_i=\frac{1+\sqrt{5}}{2})=1-\frac{1+\sqrt{5}}{2\sqrt{5}}.
\end{eqnarray*}
The bootstrap critical values are computed by 1000 bootstrap replications.
The Gaussian-based kernel of order 4, $Q(u)=(u^4-7u^2+6)\phi(u)/2$, is used to estimate the nonparametric function $g(\cdot)$, where $\phi(\cdot)$ denotes the standard normal density, see Fan and Hu (1992). For both DREAM and  Fan and Li's (1996) test, we use the Quartic kernel function as $K(u)=\frac{15}{16}(1-u^2)^2$, if $|u|\leq 1$ and $0$, otherwise, in constructing the test statistic such as that in (\ref{(2.4)}). To determine the structural dimension $q$, $c_n=0.1\log{n}/nh^{\frac{q_1}{2}}$ is used.

 The observations $\{x_i\}^n_{i=1}$ and $\{w_i\}^n_{i=1}$ are i.i.d., respectively, from multivariate normal distribution $N(0,\Sigma_1)$, $N(0,\Sigma_2)$, or $N(0,\Sigma_3)$ and independent of the standard normal errors, in which  $\Sigma_1=(\sigma_{ij}^{(1)})$, $\Sigma_2=(\sigma_{ij}^{(2)})$  and
$\Sigma_3=(\sigma_{ij}^{(3)})$ with
$\sigma_{ij}^{(1)}=I(i=j)+ 0 I(i \neq j)$,
$\sigma_{ij}^{(2)}=I(i=j)+ 0.5^{|i-j|} I(i \neq j)$ and $\sigma_{ij}^{(3)}=I(i=j)+ 0.2I(i \neq j)$.

In this section, we design 4 examples. The first example is to show that when the dimensions $p_1$ and $p_2$ are small and the model is low-frequent, how the performances of the three competitors are. In Example~2, the dimensions grow up to higher under a high-frequent model, we then check the impact from the dimensionality. Example~3 is to examine whether DREAM is still omnibus even when the test statistic  fully uses the information of low dimensionality under the null hypothesis. The model in Example~4 is with higher dimension of $B^{\top}Z$ and then we can see whether, like existing local smoothing tests, DREAM  also fails to work.  The details are in the examples.

\textbf{Example 1.} Consider the  linear regression model:
\begin{itemize}
 \item  $Y= 2\beta^{\top}_1X + 2a\times \beta^{\top}_2W + 0.5 \times \epsilon $,
 \end{itemize}
where $\beta_1=(\underbrace{1,\cdots,1}_{p_1/2},0,\cdots,0)^\top/\sqrt{p_1/2}$, $\beta_2=(0,\cdots,0,\underbrace{1,\cdots,1}_{p_2/2})^\top/\sqrt{p_2/2}$. In this example,  $p_1= p_2=2$ and $p_1= p_2=4$ are considered, where  the hypothetical and alternative models respectively respond to $a = 0$ and $ a \neq 0$. To check the sensitivity of the bandwidth selection,
we choose the different bandwidths $h=c\times n^{-1/4+\hat{q}}$ for $c=1, 1.25, 1.5, 1.75, 2$.

\begin{figure}
  \centering
\includegraphics[width=6cm]{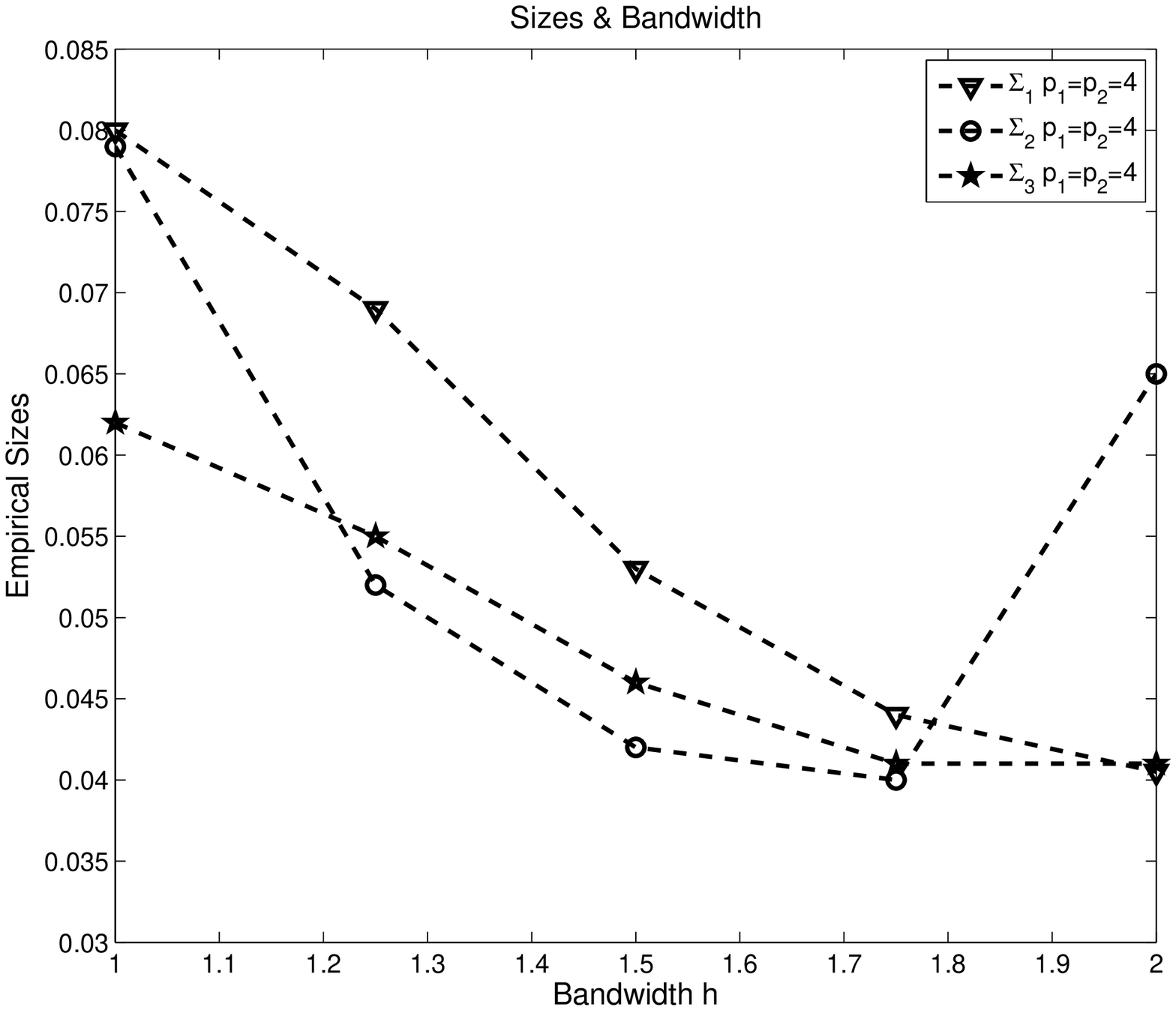}
\includegraphics[width=6cm]{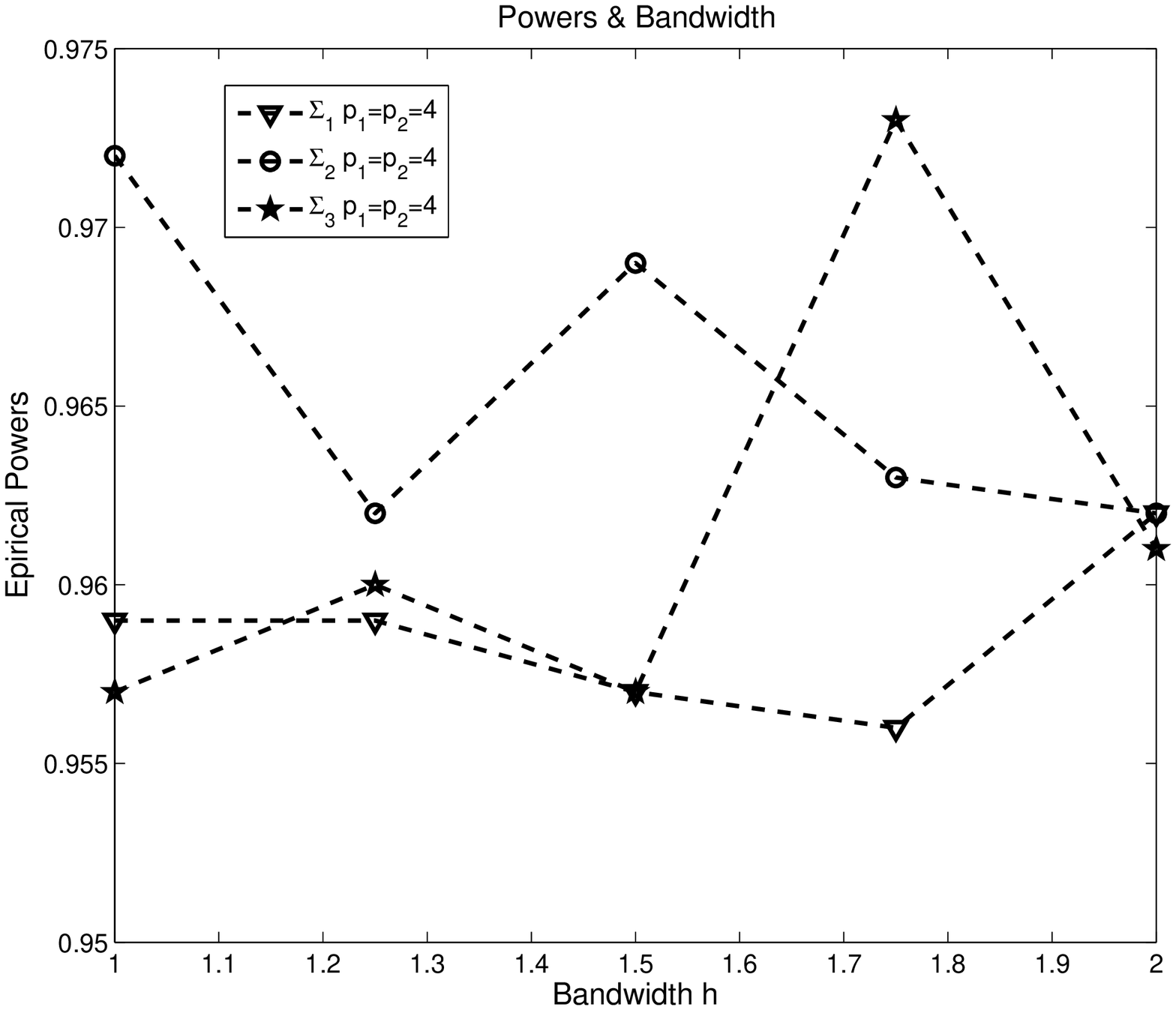}\\
  \caption{ The empirical sizes and powers curves of $T^{DEE}_{n}$ against the bandwidth $h$ with sample size 200 at the significance level $\alpha=0.05$ for Example~1.
           }\label{figure3}
\end{figure}

\begin{center}
Figure~\ref{figure3} about here
\end{center}
Figure~\ref{figure3} reports the empirical sizes and  powers with the above  bandwidths when $n = 200$.
The empirical power  is relatively robust against the different bandwidths that we use. The empirical size is not very sensitive to the bandwidth. Thus,  the bandwidth $h=1.75\times n^{-1/(4+\hat{q})}$ is recommended throughout the simulations.

The results of   the three tests under different combinations of sample sizes, dimensions of covariates $X$ and $W$ and covariance matrices $\Sigma$ are reported in Tables~\ref{table1} and \ref{table2}.
\begin{center}
Tables~\ref{table1} and \ref{table2} about here
\end{center}
\vspace{3mm}
From Table~\ref{table1}, we can clearly observe that when the dimensions $p_1$ and $p_2$ are lower, all the tests have similar empirical powers and can control the empirical sizes well. The power performances of the competitors are very good. However, from  Table~\ref{table2}, we can see that   with increasing the dimensions $p_1$ and $p_2$,   $T^{DEE}_n$ is significantly and uniformly more powerful than $T^{FL}_n$ and $T^{DM}_n$. Meanwhile $T^{DEE}_n$ can still well maintain the significance level.   Comparing Table~\ref{table1} with Table~\ref{table2}, we can see that the dimensions of $X$ and $W$ have less influence for $T^{DEE}_n$ than they do for $T^{FL}_n$ and $T^{DM}_n$.
When $p_1=p_2=4$, Fan and Li's (1996) test completely fails to detect the alternative hypothesis with a power similar to the significance level even when $n=200$. Further, DREAM is robust against the correlation structure
of $(X,Z)$ whereas  it significantly influences the power performance of Delgado and Gonz\'alez Manteiga's (2001) test, particularly when $p_1=p_2=4$.

\textbf{Example 2.} In this example, consider a nonlinear high-frequency regression model as:
\begin{itemize}
 \item  $Y= 2\sin(\beta^{\top}_1X) + 2a\times \sin(\beta^{\top}_2W) + 0.5 \times \epsilon $,
 \end{itemize}
where $\beta_1=(\underbrace{1,\cdots,1}_{p_1/2},0,\cdots,0)^\top/\sqrt{p_1/2}$ and $\beta_2=(0,\cdots,0,\underbrace{1,\cdots,1}_{p_2/2})^\top/\sqrt{p_2/2}$.
 We also consider two cases of dimensions: $p_1= p_2=4$ and $p_1= p_2 =6$. Again $a=0$ responds to the hypothetical model.
\begin{center}
Tables~\ref{table3} and \ref{table4} about here
\end{center}
\vspace{3mm}
The results are presented in Tables~\ref{table3} and \ref{table4}.
Comparing Table~\ref{table2} with Table~\ref{table3}-~\ref{table4},
we find that when the dimensions of $X$ and $W$ grow up to $p_1=p_2=6$, the empirical powers of $T^{FL}_n$ and $T^{DM}_n$ are close to 0.
This result means that both the competitors completely fail to detect the alternative hypothesis.
We can also find that the empirical power of DREAM is similar to that when $p_1=p_2=4$ in Table~\ref{table2}. This again suggests that the dimensions of $X$ and $W$ have much less impact for $T^{DEE}_n$ than they do for $T^{FL}_n$ and $T^{DM}_n$.

The next example is to confirm that DREAM is   still omnibus rather than  directional even when DREAM fully uses the information under the null hypothesis.
\textbf{Example 3} The data are generated from the following model:
\begin{itemize}
 \item  $Y= 2\sin(\beta^{\top}_1X) + \exp(\beta^{\top}_2X/2) +2a\times \sin(\beta^{\top}_2W) + 0.5 \times \epsilon $,
 \end{itemize}
where $p_1=4$, $p_2=4$, $\beta_1=(1,1,0,0)^{\top}/\sqrt{2}$ and $\beta_2=(0,0,1,1)^{\top}/\sqrt{2}$. Thus, $q_1=2,$ and $q=3.$
In this model, the conditional expectation of $E(Y|B_1^TX)$ is the same under the null and alternative hypothesis. If we simply use  this function to define a test, the alternative hypothesis cannot be detected at all from the theoretical point of view. However, DREAM can automatically adapt to the alternative model with the matrix $B^TZ$,  thus it still works under the alternative.

\begin{center}
Table~\ref{table5} about here
\end{center}
By the comparison between Tables~\ref{table2}, ~\ref{table3} and~\ref{table5}, we observe that the power performances of DREAM  in Example 3 are similar  to those in Examples 1 and 2. This means  that the test can have the advantage of dimension reduction and is still  omnibus.
The following example considers higher dimensional $B^{\top}Z$ in a model.

\textbf{Example 4} The data are generated from the following model:
\begin{itemize}
\item  $Y= X_1+0.2\exp(X_2)+a\times\frac{1.5(W_1+W_2)}{0.5+(1.5W_3+0.5)^{1.5}} +0.75\sin(W_4+1) + 0.2 \times \epsilon $,
\end{itemize}
where $p_1=4$, $p_2=4$, $X\sim N(0,4\Sigma_1)$ and $W\sim N(0,4\Sigma_1)$ and $\epsilon\sim N(0,1)$. Thus, $B_1$ is a $4\times 2$ matrix with $\beta_1=(1, 0,0,0)^{\top}$ and $\beta_2=(0,1,0,0)^{\top}$ and $B$ is a $8\times 5$ matrix with low-right block in which the columns are $b_1=(1, 1, 0,0)^{\top}$, $b_2=(0,0,1,0)^{\top}$, and $b_3=(0,0,0,1)^{\top}$. The results are summarized  in Table~\ref{table7}.
\begin{center}
Table~\ref{table7} about here
\end{center}
 Compared with the results in examples 1-3, we can see that $T^{DEE}_n$ can maintains the significance level well and its power reasonably becomes lower, but is still higher than  those of $T^{FL}_n$ and $T^{DM}_n$.  %Maybe we do not accurately determine $q$, under the alternative hypothesis, the frequency of $\hat{q}>2$ is large. This implies that
%the new estimation can automatically adapt the central mean space $\emph{S}_{E(Y|Z)}$.

In summary, the above simulations sustain the aforementioned theoretical properties that the proposed test is significantly superior to existing tests among which Fan and Li's (1996) test and Delgado and
Gonz\'alez Manteiga's (2001) test  are regarded  as  representatives of existing tests.
%The proposed test acquires a faster convergence rate to its limit under the null hypothesis and
%it can detect the sequence of the local alternative hypotheses converging the null model as a faster rate. The empirical size shows it is reasonable that the critical values can be determined by simply applying the limiting null distribution for DREAM. Hence, it is not heavy for computational burden.
%

\subsection{Baseball hitters' salary data}
We now analyze the well-known Baseball hitters' salary data set, which was originally published for the 1988 ASA Statistical Graphics and Computing
Data Exposition and is available at  \url{http://euclid.psych.yorku.ca/ftp/
sas/sssg/data/baseball.sas}. %This data set was previously analysed by Chaudhuri et al. (1994) and Xia et al. (2002).
The data set consists of information on salary and 16 performance measures of 263
major league baseball hitters. As always, the question of main interest is whether salary
reflects performance. As displayed by Friendly (2002), the 16 measures naturally belong
to three performance categories: the season hitting statistics, which include the numbers of times
at bat ($X_1$), hits ($X_2$), home runs ($X_3$), runs ($X_4$), runs batted in ($X_5$), and walks ($X_6$) in 1986; the career hitting statistics, which include the numbers of years in the major leagues ($X_7$), times at bat ($X_8$), hits ($X_9$), home runs ($X_{10}$), runs ($X_{11}$), runs batted in ($X_{12}$) and walks
($X_{13}$) during the players' entire career up to 1986; and the fielding variables, which include the
numbers of putouts ($X_{14}$), assists ($X_{15}$) and errors ($X_{16}$) in 1986.

Further, the covariates from different groups have weak correlations. The logarithm of annual salary in 1987 is used to be the response variable ($Y$) and the new covariates from the career totals by dividing totals by years in the major leagues are constructed. Let $X^{*}_j = X_j/X_7 $  for $j = 8, \cdots, 13$.
As remarked by Hoaglin and Velleman (1995), the analyses working with ln(salary) and with the
annual rate covariates fared better than those worked with the raw forms of these covariates.
Below, we  use $X^{*}_j$'s instead. All the covariates are standardized to have mean zero and unit
length. We write $V_1 = (X1, \cdots, X6)$, $V_2 = (X_7,X^{*}_8 , \cdots ,X^{*}_{13})$ and $V_3 = (X_{14}, X_{15} ,X_{16})$.
In this application, we consider two cases:
\begin{itemize}
 \item [] Case (I): $X=(V_1, V_2)$ and $W=V_3$;
 \item [] Case (II): $X=V_1$ and $W=(V_2, V_3)$;
 \end{itemize}
%\begin{center}
%Table~\ref{table8} about here
%\end{center}

Under the two cases, the values of the test statistics are respectively $1.1171$ and $6.5831$ and the corresponding $p-$values are $0.1320$ and $0.0000$.

%for different bandwidths and cases are presented in Table~\ref{table8}. Under Case (I), we can clearly observe that for different bandwidths $h$, we have the same $p-$value 0. This implies that under Case (I), we will reject the null hypothesis at the significance level $\alpha = 0.05$.  Under Case (II), applying different bandwidths $h$, we can see that the null hypothesis will be accepted at the significance level $\alpha = 0.05$, even at the significance level $\alpha = 0.10$.
%Table~\ref{table8} shows that the group $V_3$ are  the insignificance variables to the response variables under Case (II).
From these results under  Cases (I) and  (II), we can conclude that the career hitting statistic of the group $V_2$ has positive impact for the annal salary. The results are consistent with those advised by Xia et al. (2002) who found that the variables $X_{7}$, $X_{9}$ and $X_{13}$ in the group $V_2$ are  prominently to affect the annal salary. The coefficients of the fielding covariates in  the group $V_3$ are closed to $0$ in the estimated directions suggested by Xia et al. (2002).  Therefore,  for the annual salary, the group $V_2$ contains the significance covariates while the group $V_3$ does not.

\section{Conclusions}
In this paper, we  develop a dimension reduction model-adaptive test
to determine significant covariates under the nonparametric regression framework. The approach employs a dimension reduction technique to reduce the dimension such that the constructed test can well maintain the significance level and more powerful than existing  tests in the literature.
This methodology can be applicable to check other semi-parametric regression models, for example partially linear models, single-index models and partially linear single-index models.  The research is on-going. Further, as the test involves  nonparametric estimation under the null hypothesis, when the dimension of $X$ (or $B_1^{\top}X$) is high, none of tests could work well. Thus, it deserves a further study.

\section{Appendix}
\subsection{Regularity Conditions}
To prove the asymptotic properties in Sections 2 and 3, we provide the following regularity conditions:
\begin{itemize}
\item [A1] $M_n(t)$ has the following expansion:
    $$M_n(t)=M(t)+E_n\{\psi
    (Z,Y,t)\}+R_n(t),$$where $E_n(\cdot)$ denotes the average over all sample points,
    $E(\psi (Z,Y,t))=0$ and $E\{\psi^2(Z,Y,t)\}<\infty$.

\item [A2]$\sup_t ||R_n(t)||_F=o_p(n^{-1/2})$, where $||\cdot||_F$ denotes the
    Frobenius norm of a matrix.

\item [A3] The estimate $\tilde{M}_n(t)$ has the following expansion:
    $$\tilde{M}_n(t)=\tilde{M}(t)+E_n\{\tilde{\psi}
    (X,Y,t)\}+\tilde{R}_n(t),$$
  where $\tilde{M}_n({y}_i)$ is an estimate of the $ p_1 \times p_1$ positive semi-definite matrix $\tilde{M}(t)$ satisfied $\rm{Span}\{M(t)\} =\emph{S}_{\Upsilon(t)|X}$, $E(\tilde{\psi} (X,Y,t))=0$, $E\{\tilde{\psi}^2(X,Y,t)\}<\infty$ and $\sup_t ||\tilde{R}_n(t)||_F=o_p(n^{-1/2})$.
 Corresponding, $\tilde{M}_n=\frac{1}{n}\sum^{n}_{i=1}\tilde{M}_n({y}_i)$ is an estimate of the target matrix $\tilde{M}$ satisfied $\rm{Span}\{\tilde{M}\} =\emph{S}_{Y|E(Y|X)}$.

\item[A4] $(B^{\top}z_i,y_i)_{i=1}^n$ is from the probability distribution $F(B^{\top}z,y)$ on $\mathbb{R}^q \times \mathbb{R}$. The error $\varepsilon=Y-m(B^{\top}Z)$ satisfies that $E(\varepsilon^8|B^{\top}Z=B^{\top}z)$ is continuous and $E(\varepsilon^8|B^{\top}Z=B^{\top}z)\leq b(B^{\top}z)$ almost surely, where $b(B^{\top}z)$ is a measurable function such that $E(b^2(B^{\top}Z))<\infty$.

\item[A5] The density function $p_{B_1}(\cdot)$ of $B^{\top}_1X$ exists with support $\mathbb{C}$ and has a continuous and bounded first-order derivative on the support
$\mathbb{C}$. The density $p_{B_1}(\cdot)$ satisfies
    \begin{eqnarray*}
    0<\inf_{B^{\top}_1X \in \mathbb{C}} p_{B_1}(B^{\top}_1X) \leq \sup_{B^{\top}_1X \in \mathbb{C}}p_{B_1}(B^{\top}_1X) < \infty.
     \end{eqnarray*}

\item[A6] The function $g(\cdot)$ is $\eta$-order partially differentiable for some positive integer $\eta$, and the $\eta$th partially derivative of $g(\cdot)$ is bounded.

\item[A7] $\tilde{Q}(\cdot)$ is a symmetric and twice order continuously differentiable kernel function satisfying
\begin{eqnarray*}
&&\int u^{i}\tilde{Q}(u)du = \delta_{i0}, \ \ (i=0, 1,\cdots, \eta-1 ),\\
&&\tilde{Q}(u)=O[(1+|u|^{\eta+1+\epsilon})^{-1}], \quad \mbox{some}\quad \epsilon>0,
\end{eqnarray*}
 where $\delta_{ij}$ is the Kronecker's delta and $\eta$ is given in Condition A6.
\item[A8] $K(\cdot)$ is a bounded, symmetric kernel function and it is a first order continuously differentiable kernel function satisfied  $\int K(u)du = 1$.

\item[A9] $n \rightarrow \infty$, $h_1 \rightarrow 0$, $h \rightarrow 0$,
\begin{itemize}
     \item[1)] under the null or local alternative hypotheses, $nh^{q_1}_1\rightarrow \infty$, $nh^{q_1}\rightarrow \infty$ and $nh^{2\eta}_1h^{q_1/2}\rightarrow 0$;
      \item[2)] under global alternative hypothesis $H_{1}$,  $nh^{q_1}_1\rightarrow \infty$, $nh^{q}\rightarrow \infty$ and $nh^{2\eta}_1h^{q/2}\rightarrow 0$,
\end{itemize}
where $\eta$ is given in Condition A6.
\end{itemize}

\begin{remark}
Conditions A1, A2 and A3 are necessary for DEE to estimate the matrixes  $B$ and $B_1$. Under the linearity condition and constant conditional variance condition, $DEE_{SIR}$ satisfies the Conditions A1, A2 and A3.
Conditions  A4, A5, A6 and A7 are widely used for nonparametric estimation in the literature.
It is worth pointing out that Condition A6 about the higher-order kernel plays an important roles in bias reduction, see Fan and Li (1996).
Conditions A5 and A8 guarantee the asymptotic normality of DREAM statistic and make the test well-behaved.
Condition A9 about the choice of bandwidth $h$ is reasonable because the estimation $\hat{q}$ is different under the null and alternative hypotheses.
\end{remark}

\subsection{Proof of the theorems}
\textbf{Proof of Proposition~\ref{Bnnull}}.
 Note that under the null hypothesis, $E(Y|Z)=g(B^{\top}_1X)$. Then, we have
$$Y \hDash E(Y|Z)|B^{\top}_1X,$$
where the notation $\hDash$ stands for independence. This is equivalent to
 $$Y \hDash E(Y|Z)|\tilde B^{\top}Z,$$
where $\tilde B=(B_1^{\top}, O_{q_1\times p_2}^{\top})^{\top}$. From the definition of central mean subspace, $\emph{S}_{E(Y|Z)}$  is the intersection of all the  linear  spaces spanned respectively by the columns of any $d\times q$ orthogonal matrix $\Gamma$ with $1\le q\le d$ such that the above conditional independence holds. Thus,   $\emph{S}_{E(Y|Z)} \subseteq \rm{Span}(\tilde B)$ where $\rm{Span}(\tilde B)$ is the linear space spanned by the columns of $\tilde B$.
Let $\{\gamma_1,\cdots, \gamma_{q_1}\}$ the eigenvectors associated with  the nonzero eigenvalue eigenvalues of $M$.
As Zhu et al. (2010a) argued that  $\{\gamma_1,\cdots, \gamma_{q_1}\} \in \emph{S}_{E(Y|Z)}$,  we have  $\{\gamma_1,\cdots, \gamma_{q_1}\} \in \rm{Span}(\tilde B)$.
This implies that $\gamma_{j}$ for $j=1,\cdots q_1$ can be denoted as a linear combination of the columns of $\tilde B$. Thus,  for $j=1,\cdots q_1$, $\gamma_{j}$ has the similar form as $\gamma_{j}=(\tilde{\gamma}_{j},0^{\top}_{1\times p_2})^{\top}$
with $\tilde{\gamma}_{j}$ being a $p_1\times 1$ vector. This implies that any element in $\emph{S}_{E(Y|Z)}$ can also  be written as $ \gamma_{j}=(\tilde{\gamma}_{j},0^{\top}_{1\times p_2})^{\top}$. Further, the structural dimension of $\emph{S}_{E(Y|Z)}$ is smaller than or equal to $q_1$.  Further, we note that
under $H_0$,
$E(Y|Z)=E(Y|B^{\top}Z)=E(Y|B^{\top}_1X)$ and $E(Y|Z)=E(Y|X)$. Thus, $q=q_1= dim(\emph{S}_{E(Y|X)})$.

Under Conditions A1 and A2,  Theorem 2 in Zhu et al. (2010a) shows that $M_{n}-M=O_p(n^{-1/2})$.
From the arguments in Zhu and Fang (1996) and Zhu and Ng (1995), under some regularity conditions,  $\hat{\lambda}_{i}- \lambda_{i} =O_p(n^{-1/2})$, where $\hat{\lambda}_{d} \leq \hat{\lambda}_{d-1}\leq \cdots \leq \hat{\lambda}_{1}$  are the  eigenvalues of the matrix $M_n$ and $\lambda_{i}$ are the eigenvalues of the matrix $M$.
The estimate $\hat{B}$ that consists of the eigenvectors associated with the largest $q_1$ eigenvalues of $M_{n}$ is consistent to $\tilde B= (C_1 B_1^{\top}, O_{q_1\times p_2}^{\top})^{\top}$ for a $q_1\times q_1$ orthogonal matrix $C_1$. \hfill $\fbox{}$

\textbf{Proof of Proposition~\ref{prop1}.}
From Proposition~\ref{Bnnull}, $\hat{\lambda}_{i}- \lambda_{i} =O_p(n^{-1/2})$, where $\hat{\lambda}_{d} \leq \hat{\lambda}_{d-1}\leq \cdots \leq \hat{\lambda}_{1}$  are  the eigenvalues of the matrix $M_n$.

Define
\begin{eqnarray}\label{tildelambda}
\tilde{\lambda}_{i}=\frac{\lambda_{i}}{\lambda_{i}+1}, \ \mbox{for} \ 1\leq i \leq d,
\end{eqnarray}
where $\lambda_{d}= \cdots = \lambda_{d-q}=0$ and $0< \lambda_{q}\leq \cdots \leq \lambda_{1}$  are  the eigenvalues of the target  matrix $M$.

Recall the definition of $\lambda^*_{l}$ \ref{(lambda*)} in Subsection~2.3. For any $ 1< l \leq q$, since $\lambda_{l}>0$ and $\hat{\lambda}^2_{l}=\lambda^2_{l}+O_p(1/\sqrt{n})$, we have $(\lambda^*_{l})^2=(\tilde{\lambda}_{l})^2+O_p(1/\sqrt{n}).$ On the other hand,  for any $q < l \leq d$,  as $\lambda_{l}=0$ and $\hat{\lambda}^2_{l}=\lambda^2_{l}+O_p(1/n) = O_p(1/n)$, and then
$(\lambda^*_{l})^2=O_p(1/n)$.

For any $l < q$, because  $\tilde{\lambda}^2_{l}>0$ and $\tilde{\lambda}^2_{l+1}>0$, we have
\begin{eqnarray*}
\frac{(\lambda^*_{q+1})^2+c_n}{(\lambda^*_q)^2}-\frac{(\lambda^*_{l+1})^2+c_n}{(\lambda^*_l)^2}
&=&\frac{\tilde{\lambda}^2_{q+1}+c_n+O_p(1/n)}{\tilde{\lambda}^2_{q}+c_n+O_p(1/\sqrt{n})}
-\frac{\tilde{\lambda}^2_{l+1}+c_n+O_p(1/\sqrt{n})}{\tilde{\lambda}^2_{l}+c_n+O_p(1/\sqrt{n})}\\
&=&\frac{c_n+O_p(1/n)}{\tilde{\lambda}^2_{q}+c_n+O_p(1/\sqrt{n})}
-\frac{\tilde{\lambda}^2_{l+1}+c_n+O_p(1/\sqrt{n})}{\tilde{\lambda}^2_{l}+c_n+O_p(1/\sqrt{n})}.
\end{eqnarray*}
Since $c\frac{\log{n}}{n} \leq c_n \rightarrow 0$ with some fixed $c>0$, we get
\begin{eqnarray*}
\frac{(\lambda^*_{q+1})^2+c_n}{(\lambda^*_q)^2}-\frac{(\lambda^*_{l+1})^2+c_n}{(\lambda^*_l)^2}
\rightarrow  \frac{0}{\tilde{\lambda}^2_{q_1}}-\frac{\tilde{\lambda}^2_{(l+1)}}{\tilde{\lambda}^2_{l}}
=  -\frac{\tilde{\lambda}^2_{(l+1)}}{\tilde{\lambda}^2_{l}} < 0.
\end{eqnarray*}
 For any $l>q$,  $\tilde{\lambda}_{l}=0$ and $\tilde{\lambda}^2_{q} >0$, then we have
\begin{eqnarray*}
\frac{(\lambda^*_{q+1})^2+c_n}{(\lambda^*_q)^2}-\frac{(\lambda^*_{l+1})^2+c_n}{(\lambda^*_l)^2}
&=&\frac{\tilde{\lambda}^2_{q+1}+c_n+O_p(1/n)}{\tilde{\lambda}^2_{q}+c_n + O_p(1/\sqrt{n})}-\frac{\tilde{\lambda}^2_{l+1}+c_n+O_p(1/n)}{\tilde{\lambda}^2_{l}+c_n+O_p(1/n)}\\
&= & \frac{c_n+o_p(c_n)}{\tilde{\lambda}^2_{q}+c_n+O_p(1/\sqrt{n})}-\frac{c_n+o_p(c_n)}{c_n+o_p(c_n)}\\
&\rightarrow & -1 < 0.
\end{eqnarray*}
Therefor, altogether, it is concluded that $\hat{q} = q$ in probability.  \hfill $\fbox{}$

%Lastly, following the same argument as above, it is also deduced that $\hat{q}$ by (\ref{(2.7)}) is consistent to $q$. \hfill $\fbox{}$

\textbf{Proof of Theorem~\ref{the1}. } For notational convenience, denote $z_i=(x_i,w_i)$,
 $g_i = g(B^{\top}_1x_i)$, $\hat{g}_i =\hat{g}(\hat{B}^{\top}_1x_i)$,
 $u_i= y_i-g_i$, $\hat{u}_i = y_i-\hat{g}_i$, $K_{Bij} =  K(B^{\top}(z_i -z_j)/h)$, $p_i = p_{B_1i}$ and $\hat{p}_i = \hat{p}_{\hat{B}_1i}$. Throughout this Appendix, $E_i(\cdot)=E(\cdot|z_i)$.

Note that $\hat{u}_i \equiv : y_i-\hat{g}_i = u_i -(\hat{g}_i-g_i)$. We then decompose the term $V_n$ as:
\begin{eqnarray}
V_n &=& \frac{1}{n(n-1)}\sum_{i=1}^n\sum_{i \neq j}^n\frac{1}{h^{q_1}}K_{\hat{B}ij} \mu_i\mu_j
+ \frac{1}{n(n-1)}\sum_{i=1}^n\sum_{i \neq j}^n\frac{1}{h^{q_1}}K_{\hat{B}ij}(\hat{g}_i-g_i)(\hat{g}_j-g_j)\nonumber\\
&&-2\frac{1}{n(n-1)}\sum_{i=1}^n\sum_{i \neq j}^n\frac{1}{h^{q_1}}K_{\hat{B}ij}u_i(\hat{g}_j-g_j)+o_p(n^{-1}h^{-q_1/2})\nonumber\\
&\equiv:& Q_{1n}+Q_{2n}-2Q_{3n}+o_p(n^{-1}h^{-q_1/2}). \label{(6.1)}
 \end{eqnarray}
The first equality is derived by using Lemma 2 in Guo et al. (2014), where $\hat{q}=q$.
First,  consider the term $Q_{1n}$. By Taylor expansion for $Q_{1n}$ with respect to $B$, we have
\begin{eqnarray*}
Q_{1n} \equiv : Q_{11n} + Q_{12n},
  \end{eqnarray*}
where
\begin{eqnarray}
Q_{11n} &=&\frac{1}{n(n-1)}\sum_{i=1}^n\sum_{i \neq j}^n\frac{1}{h^{q_1}}K_{Bij} \mu_i\mu_j ,\label{(Q11n)}\\
Q_{12n} &=&\frac{1}{n(n-1)}\sum_{i=1}^n\sum_{i \neq j}^n\frac{1}{h^{2q_1}}K'_{\tilde{B}ij} \mu_i\mu_j(\hat{B}-B)^{\top}(z_i-z_j)\label{(Q12n)}.
  \end{eqnarray}
Here we assert $\tilde{B}=\{\tilde{B}_{ij}\}_{d \times q_1}$ with $\tilde{B}_{ij} \in  [\min\{\hat{B}_{ij}, B_{ij}\}, \max\{\hat{B}_{ij}, B_{ij}\}]$.
Let $V=B+t(\hat{B}-B)$ with a variable $t\in (-\infty, \infty )$. Define a function as
$$f(t) =\frac{1}{n(n-1)}\sum_{i=1}^n\sum_{i \neq j}^n\frac{1}{h^{q_1}}K_{Vij} \mu_i\mu_j.$$
We have $f(1)=Q_{1n}$ and $f(0)=Q_{11n}$. By an application of mean value theorem, we have
$f(1)-f(0)=f'(\tilde{t})=Q_{12n}$ with $\tilde{t} \in (0, 1)$. Thus, we can conclude that $\tilde{B}=B+\tilde{t}(\hat{B}-B)$ with $\tilde{t} \in (0, 1)$. Corresponding, we get $\tilde{B}_{ij}=B_{ij}+\tilde{t}(\hat{B}_{ij}-B_{ij})$ with $\tilde{t} \in (0, 1)$.
This affirms the assertion that $\tilde{B}=\{\tilde{B}_{ij}\}_{d \times q_1}$ with $\tilde{B}_{ij} \in  [\min\{\hat{B}_{ij}, B_{ij}\}, \max\{\hat{B}_{ij}, B_{ij}\}]$.

Because  $||\hat{B}-B||=O_p(1/\sqrt{n})$ and the first derivative  of $K_{B}(\cdot)$  with respect to $B$ is a bounded continuity function of $B$, we conclude that replacing $\tilde{B}$ by $B$ does not affect the convergence rate of $Q_{12n}$.

In the present paper, we suppose the dimension of $B^{\top} Z$ is fixed, the term $Q_{11n}$ is an $U-$statistic. Since under $H_0$, $q=q_1$, following a similar argument as that for Lemma 3.3 in Zheng (1996), it is easy to obtain:
$$nh^{q_1/2}Q_{11n} \Rightarrow  N(0,{s}),$$
% $$nh^{q_1/2}Q_{11n}\stackrel{\mathrm{d}}{\rightarrow}  N(0,{s}),$$
where
 \begin{eqnarray*}
 {s}=2\int K^2_{B}(u)du\cdot \int (\sigma^2(B^{\top}Z))^2p^2(B^{\top}Z)dB^{\top}Z
 \end{eqnarray*}
with $\sigma^2(B^{\top}z)=E(u^2|B^{\top}Z=B^{\top}z)$. We then omit the details.

We turn to discuss the term $Q_{12n}$ in \ref{(Q12n)}. Since  $E(u_i|z_i)=0$, we have $E(Q_{12n})=0$. We then calculate the second order moment of $Q_{12n}$ as follows:
\begin{eqnarray*}
E(Q^2_{12n})&=&E\big{[}\frac{1}{n(n-1)}\sum_{i=1}^n\sum_{i \neq j}^n\frac{1}{h^{2q_1}}K'_{Bij} \mu_i\mu_j(\hat{B}-B)^{\top}(z_i-z_j)\big{]}^2\\
&=&E\big{[}\frac{1}{n^2(n-1)^2}\frac{1}{h^{4q_1}}\sum_{i=1}^n\sum_{i' \neq j'}^n\sum_{i=1}^n\sum_{i' \neq j'}^nK'_{Bij}K'_{Bi'j'} \\
 &&\mu_i\mu_j\mu_{i'}\mu_{j'}(\hat{B}-B)^{\top}(z_i-z_j) (z_{i'}-z_{j'})^{\top}(\hat{B}-B)\big{]}.
\end{eqnarray*}
Since $E(u_iu_ju_{i'}u_{j'}) \neq 0$ if and only if $i = i'$, $j = j'$ or $i = j'$, $j = i'$, we have
\begin{eqnarray*}
E(Q^2_{12n})&=&\frac{2n(n-1)}{n^2(n-1)^2}\frac{1}{h^{4q_1}}E\big{[}(K'_{B12})^2 \mu^2_1\mu^2_2(\hat{B}-B)^{\top}(z_1-z_2) (z_{1}-z_{2})^{\top}(\hat{B}_{\hat{q}}-B)\big{]}\\
&=&\frac{2}{n(n-1)}\frac{1}{h^{4q_1}}E\big{[}(K'_{B12})^2 \mu^2_1\mu^2_2(\hat{B}-B)^{\top}(z_1-z_2) (z_{1}-z_{2})^{\top}(\hat{B}_{\hat{q}}-B)\big{]}\\
&=&\frac{2}{n(n-1)}\frac{1}{h^{4q_1}}\{E(u^2_1)\}^2(\hat{B}-B)^{\top}E\big{\{}(K'_{B12})^2 (z_1-z_2) (z_{1}-z_{2})^{\top}\big{\}}(\hat{B}-B).
\end{eqnarray*}
By changing variables as
$v_1=(z_1- z_2)/h$, a further computation yields
\begin{eqnarray*}
E(Q^2_{12n})&=&\frac{1}{n(n-1)}\frac{1}{h^{4q_1}}\{E(u^2_1)\}^2\int\int(K'_{B12})^2(\hat{B}-B)^{\top} (z_1-z_2)\\
&& (z_{1}-z_{2})^{\top}(\hat{B}_{\hat{q}}-B)p(B^{\top}z_1)p(B^{\top}z_2)dz_1dz_2\\
&=&\frac{1}{n(n-1)}\frac{1}{h^{q_1}}\{E(u^2_1)\}^2\int\int(K'(u))^2
(\hat{B}-B)^{\top}uu^{\top}(\hat{B}-B)\\
&&p(B^{\top}z_1)p(B^{\top}(z_1-hu))dz_1du.
\end{eqnarray*}
By taking Taylor expansion of $p(B^{\top}(z_1-hu))$ around $z_1$ and using Conditions A4, A7, A8 and A9, we have
\begin{eqnarray*}
E(Q^2_{12n}) &=&\frac{1}{n(n-1)}\frac{1}{h^{q_1}}\{E(u^2_1)\}^2\int\int(K'(u))^2
(\hat{B}-B)^{\top}uu^{\top}(\hat{B}-B)\\
&&p(B^{\top}z_1)p(B^{\top}(z_1-hu))dz_1du\\
&=&\frac{1}{n(n-1)}\frac{1}{h^{q_1}}\{E(u^2_1)\}^2\int\int(K'(u))^2
(\hat{B}-B)^{\top}uu^{\top}(\hat{B}-B)\\
&&[p^2(B^{\top}z_1)+ p(B^{\top}z_1)p'(B^{\top}z_1)h^pu]dz_1du+o_p(\frac{1}{n(n-1)})\\
&=&\frac{1}{n(n-1)}\frac{1}{h^{q_1}}\{E(u^2_1)\}^2\int\int(K'(u))^2
(\hat{B}-B)^{\top}uu^{\top}(\hat{B}-B)p^2(B^{\top}z_1)dz_1du\\
&&+\frac{1}{n(n-1)}\frac{1}{h^{q_1}}\{E(u^2_1)\}^2\int\int(K'(u))^2
(\hat{B}-B)^{\top}uu^{\top}(\hat{B}-B)\\
&&p(B^{\top}z_1)p'(B^{\top}z_1)h^{q_1}B^{\top}u)dz_1du+o_p(\frac{1}{n(n-1)})O(\frac{1}{n})\\
&=&O_p(\frac{1}{n^2(n-1)h^{q_1}})=o_p(\frac{1}{n^2}).
\end{eqnarray*}
The application of  Chebyshiev inequality leads to  $|Q_{12n}|=o_p(n^{-1}h^{-q_1/2})$.
 Using the above results for the terms $Q_{11n}$ and $Q_{12n}$,  it is deduced that
$nh^{q_{1}/2}Q_{1n}\Rightarrow N(0,s^2)$.

Now we consider the term $Q_{2n}$ in \ref{(6.1)}.  We can derive that
\begin{eqnarray}\label{Q2}
Q_{2n}&=&\frac{1}{n(n-1)}\sum_{i=1}^n\sum_{j \neq i}^n \frac{1}{h^q}K_{\hat{B}_{\hat{q}}ij}(\hat{g}_i-g_i)(\hat{g}_j-g_j)\frac{\hat{p}_{i}}{p_{i}}\frac{\hat{p}_{j}}{p_{j}} \nonumber\\
&&+\frac{1}{n(n-1)}\sum_{i=1}^n\sum_{j \neq i}^n \frac{1}{h^q}K_{\hat{B}_{\hat{q}}ij}(\hat{g}_i-g_i)(\hat{g}_j-g_j)\left(\frac{\hat{p}_{i}-p_{i}}{p_{i}}\frac{\hat{p}_{j}-p_{j}}{p_{j}}-2\frac{(\hat{p}_{i}-p_{i})\hat{p}_{j}}{p_{i}p_{j}}\right)\nonumber\\
&\equiv:& \tilde{Q}_{2n} + o_p(\tilde{Q}_{2n}).
\end{eqnarray}
Substituting the kernel estimates $\hat{g}$ and $\hat{p}$ into $\tilde{Q}_{2n}$, we have
\begin{eqnarray*}
\tilde{Q}_{2n}&=&\frac{1}{n(n-1)^3}\sum_{i=1}^n\sum_{j \neq i}^n \sum_{l\neq i}^n \sum_{k\neq j}^n \frac{1}{h^{q_1}h^{2q_1}}\frac{1}{p_{i}p_{j}}K_{\hat{B}ij}Q_{\hat{B}_1il}Q_{\hat{B}_1jk}\\
&&\times(y_l-g(B^\top_1 x_i))(y_k-g(B^\top_1 x_j)).
\end{eqnarray*}
By an application of  Taylor expansion for $\tilde{Q}_{2n}$ with respect to $B$ and $B_1$, we can have
\begin{eqnarray*}
\tilde{Q}_{2n}& \equiv &Q_{21n}+Q_{22n},
\end{eqnarray*}
where $Q_{21n}$ and $Q_{22n}$ have the following forms:
\begin{eqnarray}
Q_{21n}&=&\frac{1}{n(n-1)^3}\sum_{i=1}^n\sum_{j \neq i}^n \sum_{l\neq i}^n \sum_{k\neq j}^n \frac{1}{h^{q_1}h^{2q_1}_1}\frac{1}{p_{i}p_{j}}K_{Bij}Q_{B_1il}Q_{B_1jk}\nonumber\\
&&(y_l-g(B^{\top}_1x_i))(y_k-g(B^{\top}_1x_j)),\label{(6.3)}\\
Q_{22n}&\equiv&(\hat{B}_{1}-B_1)^{\top} Q_{221n}+(\hat{B}_{1}-B_1)^{\top}Q_{222n}+(\hat{B}-B)^{\top}Q_{223n} \label{(Q22)}
\end{eqnarray}
with $Q_{221n}$, $Q_{222n}$ and $Q_{223n}$ being the following forms:
\begin{eqnarray}
Q_{221n}&=&\frac{1}{n(n-1)^3}\sum_{i=1}^n\sum_{j \neq i}^n \sum_{l\neq i}^n\sum_{k\neq j}^n \frac{1}{h^{q_1}h^{3q_1}_1}\frac{1}{p_{i}p_{j}}K_{\tilde{B}ij}Q'_{\tilde{B}_1il}Q_{\tilde{B}_1 jk}\nonumber\\
&&(y_l-g(B^{\top}_1x_i))(y_k-g(B^{\top}_1x_j))(x_i-x_l),\label{(Q221)} \\
Q_{222n}&=&\frac{1}{n(n-1)^3}\sum_{i=1}^n\sum_{j \neq i}^n \sum_{l\neq i}^n\sum_{k\neq j}^n\frac{1}{h^{q_1}h^{3q_1}_1}\frac{1}{p_{i}p_{j}}K_{\tilde{B}ij}Q_{\tilde{B}_1il}Q'_{\tilde{B}_1 jk}\nonumber\\
&&(y_l-g(B^{\top}_1x_i))(y_k-g(B^{\top}_1x_j))(x_j-x_k), \label{(Q222)}
\end{eqnarray}
and
\begin{eqnarray}
Q_{223n}&=&\frac{1}{n(n-1)^3}\sum_{i=1}^n\sum_{j \neq i}^n \sum_{l\neq i}^n \sum_{k\neq j}^n \frac{1}{h^{2q_1}h^{2q_1}_1}\frac{1}{p_{i}p_{j}}K'_{\tilde{B}ij}Q_{\tilde{B}_1il}Q_{\tilde{B}_1 jk}\nonumber\\
&&(y_l-g(B^{\top}_1x_i))(y_k-g(B^{\top}_1x_j))(z_i-z_j). \label{(Q223)}
\end{eqnarray}
%\begin{eqnarray*}
%Q_{22n}&=&\frac{1}{n^3(n-1)}\sum_{i=1}^n\sum_{j \neq i}\sum_{k=1}^n\sum_{l=1}^n\frac{1}{h^{q_1}h^{3q_1}_1}\frac{1}{p_{i}p_{j}}K_{\tilde{B}ij}Q'_{\tilde{B}_1il}Q_{\tilde{B}_1 jk}\\
%&&(y_l-g(B^{\top}_1x_i))(y_k-g(B^{\top}_1x_j))(\hat{B}_{1}-B_{1})^{\top}(x_i-x_l)\\
%&&+\frac{1}{n^3(n-1)}\sum_{i=1}^n\sum_{j \neq i}\sum_{k=1}^n\sum_{l=1}^n\frac{1}{h^{q_1}h^{3q_1}_1}\frac{1}{p_{i}p_{j}}K_{\tilde{B}ij}Q_{\tilde{B}_1il}Q'_{\tilde{B}_1 jk}\\
%&&(y_l-g(B^{\top}_1x_i))(y_k-g(B^{\top}_1x_j))(\hat{B}_{1}-B_{1})^{\top}(x_j-x_k)\\
%&&+\frac{1}{n^3(n-1)}\sum_{i=1}^n\sum_{j \neq i}\sum_{k=1}^n\sum_{l=1}^n\frac{1}{h^{q_1}h^{3q_1}_1}\frac{1}{p_{i}p_{j}}K'_{\tilde{B}ij}Q_{\tilde{B}_1il}Q_{\tilde{B}_1 jk}\\
%&&(y_l-g(B^{\top}_1x_i))(y_k-g(B^{\top}_1x_j))(\hat{B}-B)^{\top}(z_i-z_j)\\
%&\equiv&(\hat{B}_{1}-B)^{\top} Q_{221n}+(\hat{B}_{1}-B)^{\top}Q_{222n}+(\hat{B}-B)^{\top}Q_{223n};
%\end{eqnarray*}
Here, using the similar argument as the justification for the term $Q_{1n}$, we also conclude that  $\tilde{B}=\{\tilde{B}_{ij}\}_{d \times q_1}$ with $\tilde{B}_{ij} \in  [\min\{\hat{B}_{ij}, B_{ij}\}, \max\{\hat{B}_{ij}, B_{ij}\}]$ and $\tilde{B}_1=\{\tilde{B}_{1ij}\}_{p \times q_1}$ with $\tilde{B}_{1ij} \in  [\min\{\hat{B}_{1ij}, B_{1ij}\}, \max\{\hat{B}_{1ij}, B_{1ij}\}]$.
As proved for the term $Q_{12n}$, we also assert that replacing $\tilde{B}$ and  $\tilde{B}_1$  by $B$ and $B_1$, respectively,  do not influence the convergence rate of the term $Q_{22n}$.

Similarly as the proof of Proposition~A.1 in Fan and Li (1996), when we want to finish the proof of this theorem, what we need to prove is that $E(Q^2_{21n})=o_p(n^{-2}h^{-q_1})$.
It is obvious that the calculation of  $E(Q^2_{21n})$ would be very tedious. We first prove that $E(Q_{21n})=o_p(n^{-1}h^{-q_1/2})$.

Consider two  cases with different combinations of indices $i,j,l,k$.
\begin{enumerate}
\item [Case] I:  $\mathcal{A}_1=\{i,j,l,k$ are all different from each other\}. Denote the resulting expression as $Q_{211n}$.
Under the assumption that $nh^{2{\eta}}_1h^{q_1/2}=o_p(1)$, by applying Lemma B.1, Lemmas 2 and 3 in Robinson (1988), we have
\begin{eqnarray*}
E(Q_{211n})&=&\frac{1}{h^{q_1}h^{2{q_1}}_1}E\{E_1\{\frac{1}{p_{1}}Q_{B_112}(g(B^{\top}_1x_2)-g(B^{\top}_1x_1))\}\\
&&E_3\{\frac{1}{p_{3}}Q_{B_134}(g(B^{\top}_1x_4)-g(B^{\top}_1x_3))\}K_{B13}\};\\
&\leq& C\frac{h^{2{\eta}}_1}{h^{q_1}}E[D_g(B^{\top}_1x_1)D_g(B^{\top}_1x_3)K_{B13}]\\
&=&O_p(h^{2{\eta}}_1)=o_p(n^{-1}h^{-q_1/2}).
\end{eqnarray*}
\item [Case] II: $\mathcal{A}_2=\{i,j,l,k$  take no more than three different values.\} Denote the term as $Q_{212n}$. It is easy to derive that $E(Q_{212n})=o_p(n^{-1}h^{-q_1/2})$.
\end{enumerate}
Hence, altogether, we have  $E(Q_{21n})=E(Q_{211n})+E(Q_{212n})=o_p(n^{-1}h^{-q_1/2})$.

Now we turn to compute $E(Q^2_{21n})$. It can be  decomposed as:
\begin{eqnarray*}
E(Q^2_{21n})&=&E\big{\{}\frac{1}{n(n-1)^3}\sum_{i=1}^n\sum_{j \neq i}^n \sum_{l\neq i}^n \sum_{k\neq j}^n \frac{1}{h^{q_1}h^{2q_1}_1}\frac{1}{p_{i}p_{j}}K_{Bij}Q_{B_1il}Q_{B_1jk}\nonumber\\
&&[y_l-g(B^{\top}_1x_i)][y_k-g(B^{\top}_1x_j)]\big{\}}^2\\
&=&\frac{1}{n^2(n-1)^6}\sum_{i=1}^n\sum_{j \neq i}^n \sum_{l\neq i}^n\sum_{k\neq j}^n\sum_{i'=1}^n\sum_{j' \neq i'}^n \sum_{l'\neq i'}^n\sum_{k'\neq j'}^n\frac{1}{h^{2q_1}h^{4q_1}_1}\\
&&E\big{[}\{\frac{1}{p_{i}p_{j}}K_{Bij}Q_{B_1 il}Q_{B_1 jk}
[g(B^{\top}_1x_{l})-g(B^{\top}_1x_{i})][g(B^{\top}_1x_{k})-g(B^{\top}_1x_{j})]\}\\
&&\{\frac{1}{p_{i'}p_{j'}}K_{Bi'j'}Q_{B_1i'l'}Q_{B_1j'k'}[g(B^{\top}_1x_{l'})-g(B^{\top}_1x_{i'})][g(B^{\top}_1x_{k'})-g(B^{\top}_1x_{j'})]\}\big{]}\\
&:=&LA.
\end{eqnarray*}
Firstly, when the indices $i,j,l,k$ are all different from $i',j',l',k'$, the two parts in two different braces are independent of each other. Define the related sum as $LA1$.
Applying the same argument  as that for proving  $E(Q_{21n})=o_p(n^{-1}h^{-q_1/2})$, we derive $LA1=o_p(n^{-2}h^{-q_1})$.

Secondly, consider the case where exactly one index from  $i,j,l,k$ equals one of subscripts $i',j',l',k'$. By symmetry, we only need to compute Case (i): $i=i'$; Case (ii): $i=l'$ and Case (iii)  $l=l'$. The three cases respond the related sums defined as  $LA2$, $LA3$ and $LA4$, respectively.

Under Case (i), we have
\begin{eqnarray*}
LA2&=&\frac{1}{n^2(n-1)^6h^{2q_1}h^{4q_1}_1}\sum_{i=1}^n\sum_{j \neq i}^n \sum_{l\neq i}^n \sum_{k\neq j}^n E\big{[}\{\frac{1}{p_{i}p_{j}}K_{Bij}Q_{B_1 il}Q_{B_1 jk}\\
&&[g(B^{\top}_1x_{l})-g(B^{\top}_1x_{i})][g(B^{\top}_1x_{k})-g(B^{\top}_1x_{j})]\}\\
&&\times\sum_{j' \neq i'}^n \sum_{l'\neq i'}^n \sum_{k'\neq j'}^n
\{\frac{1}{p_{i'}p_{j'}}K_{Bi'j'}Q_{B_1i'l'}Q_{B_1j'k'}\\
&&[g(B^{\top}_1x_{l'})-g(B^{\top}_1x_{i'})][g(B^{\top}_1x_{k'})-g(B^{\top}_1x_{j'})]\}\big{]}\\
&=&\frac{1}{nh^{2q_1}h^{4q_1}_1}E\big{\{}\frac{1}{p_{1}p_{2}}K_{B12}Q_{B_1 13}Q_{B_1 24}[g(B^{\top}_1x_{3})-g(B^{\top}_1x_{1})][g(B^{\top}_1x_{4})-g(B^{\top}_1x_{2})]\\
&&\frac{1}{p_{1}p_{5}}K_{B15}Q_{B_1 16}Q_{B_1 57}[g(B^{\top}_1x_{6})-g(B^{\top}_1x_{1})][g(B^{\top}_1x_{7})-g(B^{\top}_1x_{5})]\big{\}}\\
&=&\frac{1}{nh^{2q_1}h^{4q_1}_1}E\big{\{}[E_1(g(B^{\top}_1x_{3})-g(B^{\top}_1x_{1}))Q_{B_1 13}][\frac{1}{p_{1}p_{2}}E_2(g(B^{\top}_1x_{4})-g(B^{\top}_1x_{2}))Q_{B_1 24}]\\
&&\times \frac{1}{p_{1}p_{5}}[E_1(g(B^{\top}_1x_{6})-g(B^{\top}_1x_{1}))Q_{B_1 16}][E_5(g(B^{\top}_1x_{7})-g(B^{\top}_1x_{5}))Q_{B_1 57}]\big{\}}\\
&\leq & \frac{(n-1)^4h^{4\eta}_1}{n^5h^{2q_1}}E\{\frac{1}{p_{1}p_{2}p_{1}p_{5}}
D_g(B^{\top}_1x_{1})D_g(B^{\top}_1x_{2})K_{B12}D_g(B^{\top}_1x_{1})D_g(B^{\top}_1x_{5})K_{B15}\}\\
&=&O_p(h^{4\eta}_1n^{-1})=o_p(n^{-2}h^{-q_1}).
\end{eqnarray*}
For Case (ii), we have
\begin{eqnarray*}
LA3&=&\frac{1}{n^2(n-1)^6h^{2q_1}h^{4q_1}_1}\sum_{i=1}^n\sum_{j \neq i}^n \sum_{l\neq i}^n\sum_{k\neq j}^n E\big{[}\{\frac{1}{p_{i}p_{j}}K_{Bij}Q_{B_1 il}Q_{B_1 jk}\\
&&[g(B^{\top}_1x_{l})-g(B^{\top}_1x_{i})][g(B^{\top}_1x_{k})-g(B^{\top}_1x_{j})]\}\\
&&\times\sum_{i' \neq i}^n \sum_{j' \neq i'}^n \sum_{k'\neq j'}^n \{\frac{1}{p_{i'}p_{j'}}K_{Bi'j'}Q_{B_1i'i}Q_{B_1j'k'}\\
&&[g(B^{\top}_1x_{i})-g(B^{\top}_1x_{i'})][g(B^{\top}_1x_{k'})-g(B^{\top}_1x_{j'})]\}\big{]}\\
&=&\frac{1}{nh^{2q_1}h^{4q_1}_1}E\{\frac{1}{p_{1}p_{2}}K_{B12}Q_{B_1 13}Q_{B_1 24}[g(B^{\top}_1x_{3})-g(B^{\top}_1x_{1})][g(B^{\top}_1x_{4})-g(B^{\top}_1x_{2})]\\
&&\frac{1}{p_{5}p_{6}}K_{B56}Q_{B_1 15}Q_{B_1 67}[g(B^{\top}_1x_{1})-g(B^{\top}_1x_{5})][g(B^{\top}_1x_{7})-g(B^{\top}_1x_{6})]\}\\
&=&\frac{1}{nh^{2q_1}h^{4q_1}_1}E\big{\{}K_{B12}[E_1(g(B^{\top}_1x_{3})-g(B^{\top}_1x_{1}))Q_{B_1 13}][\frac{1}{p_{1}p_{2}}E_2(g(B^{\top}_1x_{4})-g(B^{\top}_1x_{2}))Q_{B_1 24}]\\
&&\times \frac{1}{p_{5}p_{6}}K_{B56}E_1[(g(B^{\top}_1x_{1})-g(B^{\top}_1x_{5}))Q_{B_1 15}][E_6(g(B^{\top}_1x_{7})-g(B^{\top}_1x_{6}))Q_{B_1 67}]\big{\}}\\
&\leq & \frac{(n-1)^4h^{4\eta}_1}{n^5h^{2q_1}}E\{\frac{1}{p_{1}p_{2}p_{1}p_{5}}D_g(B^{\top}_1x_{1})D_g(B^{\top}_1x_{2})K_{B12}D_g(B^{\top}_1x_{1})D_g(B^{\top}_1x_{6})K_{B56}\}\\
&=&O_p(h^{3\eta}_1n^{-1})=o_p(n^{-2}h^{-q_1}).
\end{eqnarray*}
Similarly, it is easy to prove that for Case (iii), $LA4=o_p(n^{-2}h^{-q_1})$.

Finally, when the indices $i,j,l,k,i', j', l', k'$ take no more than six different values whose sum is defined as $LA5$, it is easy to see that $LA5=o_p(n^{-2}h^{-q_1})$.

Combining all cases, we get $E(Q^2_{21n})= LA1+LA2 +9LA3+6LA4+LA5=o_p(n^{-2}h^{-q_1})$.  The application of Chebyshiev's inequality yields $Q_{21n}=o_p(n^{-1}h^{-q_1/2})$.

The similar arguments can be applied to handle the first and second moment of  $Q_{22n}$ defined in \ref{(Q22)}. The first moment is very similar to that for $Q_{21n}$. Note that $Q_{22n}$ is the weighted sum of $Q_{221n}$, $Q_{222n}$ and  $Q_{223n}$ in \ref{(Q221)}--\ref{(Q223)}. Thus, we only need to respectively handle $Q_{221n}$, $Q_{222n}$ and  $Q_{223n}$. Also, they can be treated similarly as those for $Q_{21n}$. When Conditions~A4$-$A9 hold, it is easy to derive the following convergence rates:
\begin{eqnarray}
E(Q^2_{221n})&=&O_p(\max\{h^{2\eta}_1, h^{2\eta}_1n^{-1}, h^{\eta}_1n^{-1}\})=o_p(n^{-1}h^{-q_1/2}),\label{Result1} \\
 E(Q^2_{222n})&=&O_p(\max\{h^{2\eta}_1, h^{2\eta}_1n^{-1}, h^{\eta}_1n^{-1}\})=o_p(n^{-1}h^{-q_1/2}),\label{Result2}\\
 E(Q^2_{223n})&=&O_p(\max\{h^{4\eta}_1, h^{4\eta}_1n^{-1}, h^{3\eta}_1n^{-1}\})=o_p(n^{-1}h^{-q_1/2}).\label{Result3}
\end{eqnarray}
Due to the facts $||\hat{B}-B||=O_p(1/\sqrt{n})$ and  $||\hat{B}_1-B_1||=O_p(1/\sqrt{n})$, together with \ref{Result1}--\ref{Result3}, we derive $Q_{22n}=o_p(n^{-1}h^{-q_1/2})$ by an application of Chebyshiev's inequality.
Therefore, altogether, from the definition of $Q_{2n}$ in (\ref{Q2}), we have that $Q_{2n}=o_p(n^{-1}h^{-q_1/2})$.

Lastly, we consider the terms $Q_{3n}$ in \ref{(6.1)}. Also it is easy to see that
\begin{eqnarray*}
Q_{3n} &=& \frac{1}{n(n-1)}\sum_{i=1}^n\sum_{i \neq j}^n \frac{1}{h^{q_1}}K_{\hat{B}ij}u_i(\hat{g}_j-g_j)\frac{\hat{p}_{j}}{p_{j}}\\
&&+\frac{1}{n(n-1)}\sum_{i=1}^n\sum_{i \neq j}^n \frac{1}{h^{q_1}}K_{\hat{B}ij}u_i(\hat{g}_j-g_j)\left(\frac{\hat{p}_{j}-p_{j}}{p_{j}}\right)\\
&\equiv:& \tilde{Q}_{3n} + o_p(\tilde{Q}_{3n}),
 \end{eqnarray*}
substituting the kernel estimates $\hat{g}$ and $\hat{p}$ into $\tilde{Q}_{3n}$, we have
\begin{eqnarray*}
\tilde{Q}_{3n}&=&\frac{1}{n^2(n-1)}\sum_{i=1}^n\sum_{j \neq i}^n\sum_{k\neq j}^n \frac{1}{h^{q_1}h^{q_1}_1}\frac{1}{p_{j}}K_{\hat{B}ij}u_iQ_{\hat{B}_1jk}(y_k-g(B^\top_1 x_j)).
\end{eqnarray*}
By using the Taylor expansion for $\tilde{Q}_{3n}$ with respect to $B$ and $B_1$, we can have
\begin{eqnarray*}
\tilde{Q}_{3n} \equiv: Q_{31n}+Q_{32n},
\end{eqnarray*}
where $Q_{31n}$ and $Q_{32n}$  have following forms:
\begin{eqnarray*}
Q_{31n}&=&\frac{1}{n(n-1)^2}\sum_{i=1}^n\sum_{j \neq i}^n\sum_{k\neq j}^n \frac{1}{h^{q_1}h^{q_1}_1}\frac{1}{p_{j}}K_{Bij}u_iQ_{B_1jk}(y_k-g(B^\top_1 x_j))
\end{eqnarray*}
and
\begin{eqnarray*}
Q_{32n}&=&\frac{1}{n(n-1)^2}\sum_{i=1}^n\sum_{j \neq i}^n \sum_{k\neq j}^n \frac{1}{h^{q_1}h^{q_1}_1}\frac{1}{p_{j}}K_{\tilde{B}ij}u_iQ'_{\tilde{B}_1jk}\\
&&(y_k-g(B^\top_1 x_j))(\hat{B}_{1}-B_{1})^{\top}(x_i-x_l)\\
&&+\frac{1}{n(n-1)^2}\sum_{i=1}^n\sum_{j \neq i}^n \sum_{k\neq j}^n \frac{1}{h^{q_1}h^{q_1}_1}\frac{1}{p_{j}}K'_{\tilde{B}ij}u_iQ_{\tilde{B}_1jk}\\
&&(y_k-g(B^\top_1 x_j))(\hat{B}-B)^{\top}(z_i-z_j)\\
&\equiv&(\hat{B}_{1}-B_1)^{\top} Q_{321n}+(\hat{B}-B)^{\top}Q_{322n}.
\end{eqnarray*}
Here, applying the similar justification for the term $Q_{1n}$, we also conclude   $\tilde{B}=\{\tilde{B}_{ij}\}_{d \times q_1}$ with $\tilde{B}_{ij} \in  [\min\{\hat{B}_{ij}, B_{ij}\}, \max\{\hat{B}_{ij}, B_{ij}\}]$ and $\tilde{B}_1=\{\tilde{B}_{1ij}\}_{p \times q_1}$ with $\tilde{B}_{1ij} \in  [\min\{\hat{B}_{1ij}, B_{1ij}\}, \max\{\hat{B}_{1ij}, B_{1ij}\}]$.
%where $\tilde{B}$ is within the region  $[\min\{\hat{B}, B\}, \max\{\hat{B}, B\}]$ and $\tilde{B}_1$ is within the region  $[\min\{\hat{B}_{1}, B\}, \max\{\hat{B}_{1}, B\}]$.
Similarly,  replacing  $\tilde{B}$ and  $\tilde{B}_1$  by $B$ and $B_1$, respectively, does not effect the convergence rate of the term $Q_{32n}$.

Because $E(u_i|z_i)=0$, we have $E(Q_{31n})=0$. Then we compute the second  moment of $Q_{31n}$ as follows:
\begin{eqnarray*}
E(Q^2_{31n})&=&E\big{[}\frac{1}{n(n-1)^2}\sum_{i=1}^n\sum_{j \neq i}^n\sum_{k\neq j}^n \frac{1}{h^{q_1}h^{q_1}_1}\frac{1}{p_{j}}K_{Bij}u_iQ_{B_1jk}(y_k-g(B^\top_1 x_j))\big{]}^2\\
&=&E\big{[}\frac{1}{n^2(n-1)^4}\frac{1}{h^{2q_1}h^{2q_1}_1}\sum_{i=1}^n\sum_{j \neq i}^n\sum_{k\neq j}^n\sum_{i'=1}^n\sum_{j' \neq i'}^n\sum_{k'\neq j'}^n
\frac{1}{p_{j}p_{j'}}K_{Bij}Q_{B_1jk}K_{Bi'j'}Q_{B_1j'k'} \\
&&u_iu_{i'}(y_k-g(B^\top_1 x_j))(y_k'-g(B^\top_1 x_j'))\big{]}\\
&=&E\big{[}\frac{1}{n^2(n-1)^4}\frac{1}{h^{2q_1}h^{2q_1}_1}\sum_{i=1}^n\sum_{j \neq i}^n\sum_{k\neq j}^n \sum_{i'=1}^n\sum_{j' \neq i'}^n\sum_{k'\neq j'}^n \frac{1}{p_{j}p_{j'}}K_{Bij}Q_{B_1jk}K_{Bi'j'}Q_{B_1j'k'} \\
&&u_iu_{i'}(g(B^\top_1 x_k)-g(B^\top_1 x_j))(g(B^\top_1 x_{k'})-g(B^\top_1 x_j'))\big{]}+o_p(n^{-2}h^{-q_1}).\\
\end{eqnarray*}
Since $E(u_iu_{i'}) \neq 0$ if and only if $i = i'$, we have
\begin{eqnarray*}
E(Q^2_{31n})&=&\frac{1}{n}\frac{1}{h^{2q_1}h^{2q_1}_1}E(u^2_1)E\big{[}\frac{1}{p_{2}p_{4}}K_{B12}Q_{B_123}K_{B14}Q_{B_145} \\
&&(g(B^\top_1 x_3)-g(B^\top_1 x_2))(g(B^\top_1 x_5)-g(B^\top_1 x_4))\big{]}\\
&=&\frac{1}{n}\frac{1}{h^{2q_1}h^{2q_1}_1}E(u^2_1)E\big{[}\frac{1}{p_{2}p_{4}}K_{B12}K_{B14}E_2[Q_{B_123}(g(B^\top_1 x_3)-g(B^\top_1 x_2))]\\
&&\times E_4[Q_{B_145}(g(B^\top_1 x_5)-g(B^\top_1 x_4))]\big{]}\\
&\leq&\frac{1}{n}\frac{h^{2\eta}_1}{h^{2q_1}}E\{\frac{1}{p_{2}p_{4}}K_{B12}K_{B14}D_g(B^{\top}_1x_{2})D_g(B^{\top}_1x_{4})\}\\
&=&O_p(h^{2\eta}_1n^{-1})=o_p(n^{-2}h^{-q_1}),
\end{eqnarray*}
Employing Lemma B.1, Lemmas 2 and 3 in Robinson (1988) again, we have $Q_{31n}=o_p(n^{-1}h^{-q_1/2})$. Also, following the similar arguments used for proving $Q_{22n}$ %as the proof of the term $Q_{31n}$, since  $||\hat{B}-B||=O_p(1/\sqrt{n})$ and  $||\hat{B}_1-B_1||=O_p(1/\sqrt{n})$,
we get $Q_{32n}=o_p(n^{-1}h^{-q_1/2})$ and then $Q_{3n}=o_p(n^{-1}h^{-q_1/2})$.

In summary, we conclude that:
$$n h^{q_1/2}V_n {\Rightarrow}  N(0, s^2).$$
%\begin{eqnarray*}
%n h^{q_1/2}V_n \stackrel{\mathrm{d}}{\rightarrow}  N(0, s^2).
%\end{eqnarray*}
 the variance $s^2$ can be estimated by
  \begin{eqnarray*}
\hat{s}^2 = \frac{2}{n(n-1)}\sum_{i=1}^n\sum_{j \neq i}^nK^2_h\left(\hat{B}^{\top}z_i-\hat{B}^{\top}z_j\right)\hat{u}^2_i\hat{u}^2_j.
\end{eqnarray*}
Since the proof is  straightforward, we only give a very
brief outline. Under the null hypothesis,  the estimates $\hat{B}$ and $\hat{g}$ are consistent to $B$ and $g$, and  some elementary calculations result in an asymptotic presentation as:
$$\hat{s}^2 = \frac{2}{n(n-1)}\sum_{i=1}^n\sum_{j \neq i}^nK^2_h\left(\hat{B}^{\top}z_i-\hat{B}^{\top}z_j\right)u^2_iu^2_j+o_p(1).$$
Applying a similar argument used for proving Lemma~2 in Guo et al. (2014), one can derive
\begin{eqnarray*}
\hat{s}^2 &=& \frac{2}{n(n-1)}\sum_{i=1}^n\sum_{j \neq i}^nK^2_h\left(B^{\top}z_i-B^{\top}z_j\right)u^2_iu^2_j+o_p(1)\\
&\equiv:& \tilde{s}^2+o_p(1).
\end{eqnarray*}
As $\tilde{s}^2$ is an U-statistic and it is easy to prove $\tilde{s}^2 \rightarrow s^2$ in probability.  The more details can be referred to Zheng (1996). The proof of Theorem~\ref{the1} is finished. \hfill $\fbox{}$

\textbf{Proof of Lemma~\ref{lemma1}}. %Here we adopt the similar justification of Lemma 3.1 in Zhu et al.  (2015a).
Consider  RERE that is based on $DEE_{SIR}$. From the proof of Theorem~3.2 in Li et al. (2008), we see that to detain $M_{n}-M = O_p(C_n)$, it is only needed to prove $M_n(t)- M(t) = O_p(C_n)$ uniformly,  where
$M(t) = \Sigma^{-1}Var(E(Z|I(Y\leq t))=\Sigma^{-1}(\nu_1-\nu_0)(\nu_1-\nu_0)^{\top}p_t(1-p_t)$, $\Sigma$ is the covariance matrix of $Z$, $\nu_0 = E(Z|I(Y\leq t)=0)$, $\nu_1 = E(Z|I(Y\leq t)=1)$ and $p_t=E(I(Y\leq t))$.

Further, we note that
\begin{eqnarray*}
\nu_{1}-\nu_{0} &=&\frac{E(ZI(Y \leq t))}{p_t}-\frac{E(ZI(Y > t))}{1-p_t}\\
&=& \frac{E(ZI(Y \leq t)-E(Z))E(I(Y \leq t))}{p_t(1-p_t)}.
\end{eqnarray*}
Therefore, the matrix $M(t)$ can also be reformulated as
\begin{eqnarray*}
M(t)  &=& \Sigma^{-1}[E\{(Z-E(Z))I(Y \leq t)\}][E\{(Z-E(Z))I(Y \leq t)\}]^\top\\
&=:&\Sigma^{-1} \tilde{m}(t)\tilde{m}(t)^{\top},
\end{eqnarray*}
where $ \tilde{m}(t)= E\{(Z-E(Z))I(Y \leq t)\}$. Correspondingly,  $\tilde{m}(t)$ can be simply  estimated by
\begin{eqnarray*}
\tilde{m}_{n}(t) = n^{-1}\sum_{i=1}^n (z_{i}- \bar{z})I(y_{i} \leq t).
\end{eqnarray*}
Then $M(t)$ can be estimated by
$$M_{n}(t) =\hat{\Sigma}^{-1}L_{n}(t), $$
where $ \bar{z}=\frac{1}{n}\sum_{i=1}^nz_i$, $L_{n}(t)=\tilde{m}_{n}(t)\tilde{m}_{n}(t)^{\top}$ and $\hat{\Sigma}$ is the estimate of $\Sigma$.
Denote respectively the response under the null and local alternative hypotheses as $Y$ and $Y_n$ to show the dependence of the response under the local alternative.
Then under $H_{1n}$,
\begin{eqnarray*}
E\{ZI(Y_{n} \leq t)\}-E\{ZI(Y \leq t)\} =E[Z\{P(Y_{n} \leq t|Z)\}]-E[Z\{P(Y \leq t|Z)\}],
\end{eqnarray*}
where $Y_n=g(B^{\top}_1X)+ C_n G(B^{\top}Z)+\varepsilon \equiv: Y + C_n G(B^{\top}Z)$. Thus, for all $t$, we have
\begin{eqnarray*}
P(Y_{n} \leq t|Z)-P(Y \leq t|Z)
&=&F_{Y|Z}(t-C_nG(B^{\top}Z))-F_{Y|Z}(t)\\
&=&-C_nG(B^{\top}Z)f_{Y|Z}(t)+O_p(C_n).
\end{eqnarray*}
Under Condition A2, we can conclude that $n^{-1}\sum_{i=1}^n z_{i}I(y_{ni} \leq t)-E\{ZI(Y \leq t)\}=O_p(\max{(C_n, n^{-1/2})}) = O_p(C_n)$. By the parallel argument for  justifying Theorem 3.2 of Li et al. (2008),  $M_n(t)-M(t) = O_p(C_n)$ uniformly and then  $M_{n}-M = O_p(C_n)$.

Similarly as those in Zhu and Fang (1996) and Zhu and Ng (1995),  we get that $\hat{\lambda}_{i}- \lambda_{i} = O_p(C_n)$, where $\hat{\lambda}_{d} \leq \hat{\lambda}_{d-1}\leq \cdots \leq \hat{\lambda}_{1}$ are the eigenvalues of the matrix $M_n$. %Thus, we can use the similar statements as those in Proposition ~\ref{prop1}.
Note that under $H_0$, $\lambda_{d}= \cdots = \lambda_{q}=0$ and $0< \lambda_{q}\leq \cdots \leq \lambda_{1}$ that are the eigenvalues of the matrix $M$. Since $c C^2_n\log{n} \leq c_n \rightarrow 0$ with some fixed $c>0$ and $C_n = 1/(n^{1/2}h^{q_1/4})$, we have $C^2_n=o_p(c_n)$.
It is clear that for any $l>q$, $\lambda_{l}=0$, so we get $(\lambda^*_{l})^2=O_p(C^2_n)$ where $\lambda^*$ is defined in (\ref{(lambda*)}).
Recall the definition of $\tilde{\lambda}$ \ref{tildelambda} in the beginning of the proof of Proposition~\ref{prop1}. For any $1 \leq l \leq q$, we have   $(\lambda^*_{l})^2=(\tilde{\lambda}_{l})^2+O_p(C_n)$.
Thus, when $l>q$, we have,
\begin{eqnarray*}
\frac{(\lambda^*_{q+1})^2+c_n}{(\lambda^*_q)^2}-\frac{(\lambda^*_{l+1})^2+c_n}{(\lambda^*_l)^2}
&=&\frac{\tilde{\lambda}^2_{q+1}+c_n+O_p(C^2_n)}{\tilde{\lambda}^2_{q}+c_n + O_p(C_n)}-\frac{\tilde{\lambda}^2_{l+1}+c_n+O_p(C^2_n)}{\tilde{\lambda}^2_{l}+c_n+O_p(C^2_n)}\\
&=&\frac{\tilde{\lambda}^2_{q+1}+c_n+o_p(c_n)}{\tilde{\lambda}^2_{q}+c_n + O_p(C^2_n)}-\frac{\tilde{\lambda}^2_{l+1}+c_n+o_p(c_n)}{\tilde{\lambda}^2_{l}+c_n+o_p(c_n)}\\
&=&\frac{c_n+o_p(c_n)}{\tilde{\lambda}^2_{q}+c_n + O_p(C^2_n)}-\frac{c_n+o_p(c_n)}{c_n+o_p(c_n)}.
\end{eqnarray*}
Therefore in probability
\begin{eqnarray*}
\frac{(\lambda^*_{q+1})^2+c_n}{(\lambda^*_q)^2}-\frac{(\lambda^*_{l+1})^2+c_n}{(\lambda^*_l)^2}
\rightarrow  -1 <0.
\end{eqnarray*}
When $1 \leq l < q$, we derive that:
\begin{eqnarray*}
\frac{(\lambda^*_{q+1})^2+c_n}{(\lambda^*_q)^2}-\frac{(\lambda^*_{l+1})^2+c_n}{(\lambda^*_l)^2}
&=&\frac{\tilde{\lambda}^2_{q+1}+c_n+O_p(C^2_n)}{\tilde{\lambda}^2_{q}+c_n + O_p(C_n)}-\frac{\tilde{\lambda}^2_{l+1}+c_n+O_p(C_n)}{\tilde{\lambda}^2_{l}+c_n+O_p(C_n)}\\
&=&\frac{c_n+o_p(c_n)}{\tilde{\lambda}^2_{q}+c_n + O_p(C^2_n)}-\frac{\tilde{\lambda}^2_{l+1}+c_n+o_p(c_n)}{\tilde{\lambda}^2_{l}+c_n+o_p(c_n)}.
\end{eqnarray*}
Then we have in probability
\begin{eqnarray*}
\frac{(\lambda^*_{q+1})^2+c_n}{(\lambda^*_q)^2}-\frac{(\lambda^*_{l+1})^2+c_n}{(\lambda^*_l)^2}
\rightarrow -\frac{\tilde{\lambda}^2_{l+1}}{\tilde{\lambda}^2_{l}} <0.
\end{eqnarray*}
Therefore, altogether, we can conclude that $\hat{q} = q$ with a probability going to 1.

%Lastly, to prove that $\hat{q}$  given by (\ref{(2.7)}) tends to $q_1$, we use the same argument as the justification that $\hat{q}_1$  given by (\ref{(2.7)}) tends to $q_1$.\hfill$\Box$

\textbf{Proof of Theorem~\ref{the2}.} Prove Part (I). Since the details of the proof is similar to that of the proof of Theorem~\ref{the1}, we  only sketch it.
By using $||\hat{B}_1-B_1||=O_p(1/\sqrt{n})$ and Conditions A6 and A8 in Appendix,
$\hat{g}(\hat{B}^{\top}_1x)$ is an uniformly consistent estimate of $g(B^{\top}_1x)=E(Y|B^{\top}_1X=B^{\top}_1x)$, see Powell et al. (1989) or Robinson (1988). We then have
\begin{eqnarray*}
V_n &=& \frac{1}{n(n-1)}\sum_{i=1}^n\sum_{i \neq j}\frac{1}{h^{\hat{q}}}K_{\hat{B}ij} \hat{u}_i \hat{u}_j\\
&=& \frac{1}{n(n-1)}\sum_{i=1}^n\sum_{i \neq j}\frac{1}{h^q}K_{Bij} u_iu_j+o_p(1),
 \end{eqnarray*}
where $u_i=y_i-g(B^{\top}_1x_i)$ with $g(B^{\top}_1x_i)=E(y_i|x_i)$. Let $\triangle(z_i)=m(B^{\top}z_i)-g(B^{\top}_1x_i)$.
Therefore, by using the $U-$statistics theory, we get that
\begin{eqnarray*}
V_n = E\{K_{B12} u_1u_2\}+o_p(1)
= E\{\triangle^2(Z)p(B^{\top}Z)\}+o_p(1).
\end{eqnarray*}
 Similarly, we
can also prove that in probability $\hat{s}$ converges to a positive value which
may be different from $s$ defined by \ref{(3.1)}.
Therefore, we can obtain $T_{n}/(n h^{q_1/2}) \Rightarrow {constant} >0$ in probability.

Consider Part (II). Following the similar arguments used to prove Theorem \ref{the1}, we can show that:
\begin{eqnarray*}
V_n &=& \frac{1}{n(n-1)}\sum_{i=1}^n\sum_{i \neq j}\frac{1}{h^{\hat{q}}}K_{\hat{B}ij} \hat{u}_i \hat{u}_j\\
&=& \frac{1}{n(n-1)}\sum_{i=1}^n\sum_{i \neq j}\frac{1}{h^q}K_{\tilde{B}ij} u_iu_j+o_p((nh^{q_1})^{-1})
=: Q_{n}+o_p((nh^{q_1})^{-1})
 \end{eqnarray*}
where $u_i=y_i- g(B^{\top}_1x_i)=C_nG(B^{\top}z_i)+\varepsilon_i$
Then under $H_{1n}$, $E(\varepsilon_i|z_i)=0$. $Q_{n}$ is further decomposed as:
\begin{eqnarray*}
Q_{n}&=&\{\frac{1}{n(n-1)}\sum_{i=1}^n\sum_{j \neq i}\frac{1}{h^{q_1}}K_{\tilde{B}ij}(C_nG(B^{\top}z_i)+\varepsilon_i)(C_nG(B^{\top}z_j)+\varepsilon_j)\}\\
&=& \{\frac{1}{n(n-1)}\sum_{i=1}^n\sum_{j \neq i}\frac{1}{h^{q_1}}K_{\tilde{B}ij}\varepsilon_i\varepsilon_j\}
+C_n\{\frac{1}{n(n-1)}\sum_{i=1}^n\sum_{j \neq i}\frac{1}{h^{q_1}}K_{\tilde{B}ij}G(B^{\top}z_i)\varepsilon_{j}\}\\
&&+C^2_n\{\frac{1}{n(n-1)}\sum_{i=1}^n\sum_{j \neq i}\frac{1}{h^{q_1}}K_{\tilde{B}ij}G(B^{\top}z_i)G(B^{\top}z_j)\}\\
&\equiv& W_{1n}+C_nW_{2n}+C^2_nW_{3n},
\end{eqnarray*}
where $\tilde{B} = (B^{\top}_1, O^{\top}_{p_2 \times q_1})^{\top}$.
Again following the similar argument as that for Lemma 3.3 in Zheng (1996), we can easily derive that
$nh^{\frac{q_1}{2}}W_{1n}\stackrel{\mathrm{d}}{\rightarrow} N(0,s^2)$. By Lemma 3.1 of Zheng (1996), we get that $\sqrt{n}W_{2n}=O_p(1)$.
Since $C_n = n^{-\frac{1}{2}}h^{-\frac{q_1}{4}}$, it is deduced that $nh^{\frac{q_1}{2}}C_nW_{2n}=o_p(1)$. Lastly, we consider the term $W_{3n}$. It is obvious that the term $W_{3n}$ is $U-$statistic with the kernel as:
\begin{eqnarray*}
H(z_i, z_j)=\frac{1}{h^{q_1}}K_{\tilde{B}ij}G(B^{\top}z_i)G(B^{\top}z_j).
\end{eqnarray*}
We firstly calculate the expectation of $H(z_i, z_j)$ as
\begin{eqnarray*}
E\{H(z_i, z_j)\}&=&\frac{1}{h^{q_1}}E\{K_{\tilde{B}ij}G(B^{\top}z_i)G(B^{\top}z_j)\}\\
&=&\frac{1}{h^{q_1}}E\big[K_{\tilde{B}ij}E\{G(B^{\top}z_i)|\tilde{B}^{\top}z_i\}E\{G(B^{\top}z_j)|\tilde{B}^{\top}z_j\}\big].
\end{eqnarray*}
In order to conveniently write, suppose $M(\tilde{B}^{\top}z_i)=E\{G(B^{\top}z_i)|\tilde{B}^{\top}z_i\}$ and $t_i=\tilde{B}^{\top}z_i$. The expectation of $H(z_i, z_j)$ can be further  calculated as
\begin{eqnarray*}
E\{H(z_i, z_j)\}
&=&\frac{1}{h^{q_1}}E\big[K(\frac{t_i-t_j}{h})M(t_i)M(t_j)\big]\\
&=&\int \int\frac{1}{h^{q_1}}K(\frac{t_i-t_j}{h})M(t_i)M(t_j)p_1(t_i)p_1(t_j)dt_idt_j.
\end{eqnarray*}
%\begin{eqnarray*}
%E\{H(z_i, z_j)\}
%&=&\frac{1}{h^{q_1}}E\big[K(\frac{t_i-t_j}{h})M(t_i)M(t_j)\big]\\
%&=&\int \int\frac{1}{h^{q_1}}K(\frac{t_i-t_j}{h})M(t_i)M(t_j)p_1(t_i)p_1(t_j)dt_idt_j\\
%&=&\int \int K(u)M(t_i)M(t_i-hu)p_1(t_i)p_1(t_i-hu)dudt_i\\
%&=&\int K(u)du \int M^2(t_i)p^2_1(t_i)dt_i+o_p(h)\\
%&=&E\{M^2(t_i)p_1(t_i)\}+o_p(1)\\
%&=&E\big([E\{G(B^{\top}Z)|\tilde{B}^{\top}Z\}]^2p_{\tilde{B}}(\tilde{B}^{\top}Z)\big)+o_p(1)\\
%&=&E\big([E\{G(B^{\top}Z)|B_1^{\top}X\}]^2p_{B_1}(B_1^{\top}X)\big)+o_p(1).
%\end{eqnarray*}
Further, apply the changing variables $u=\frac{t_i-t_j}{h}$ to get that:
\begin{eqnarray*}
E\{H(z_i, z_j)\}
&=&\int \int K(u)M(t_i)M(t_i-hu)p_1(t_i)p_1(t_i-hu)dudt_i\\
&=&\int K(u)du \int M^2(t_i)p^2_1(t_i)dt_i+o_p(h)\\
&=&E\big([E\{G(B^{\top}Z)|\tilde{B}^{\top}Z\}]^2p_{\tilde{B}}(\tilde{B}^{\top}Z)\big)+o_p(1)\\
&=&E\big([E\{G(B^{\top}Z)|B_1^{\top}X\}]^2p_{B_1}(B_1^{\top}X)\big)+o_p(1).
\end{eqnarray*}
Again using the element characteristics of $U-$statistic, we derive that
$W_{3n}\Rightarrow E\big([E\{G(B^{\top}Z)|B_1^{\top}X\}]^2p_{B_1}(B_1^{\top}X)\big)$.
%$W_{3n}\stackrel{\mathrm{d}}{\rightarrow}
%E\big([E\{G(B^{\top}Z)|B_1^{\top}X\}]^2p_{B_1}(B_1^{\top}X)\big)$.

Thus, we can deduce that
$$V_{n} \Rightarrow N(E\big([E\{G(B^{\top}Z)|B_1^{\top}X\}]^2p_{B_1}(B_1^{\top}X)\big), s^2).$$ As the variance $s^2$ can be consistently estimated and thus, the expected result can be proved.

In summary, invoking Slutsky theorem, it is concluded that
$$T_{n}\Rightarrow N(E\big([E\{G(B^{\top}Z)|B_1^{\top}X\}]^2p_{B_1}(B_1^{\top}X)\big)/s, 1).$$
The proof of Theorem~\ref{the2} is concluded.  \hfill$\Box$.

\

%\begin{figure}
%  \centering
%  % Requires \usepackage{graphicx}
%  \includegraphics[width=\textwidth]{introductionfan05.eps} \\
%  \caption{ The empirical power curve of Fan and Li's (1996) test against the dimensions of $X$ and $W$ with sample size 200
%  at the significance level $\alpha=0.05$.
%           }\label{figure1}
%\end{figure}
%
%\begin{figure}
%  \centering
%  % Requires \usepackage{graphicx}
%  \includegraphics[width=\textwidth]{introductionDM05.eps} \\
%  \caption{ The empirical power curve of
%  Delgado and Manteiga's (2001) test against the dimensions of $X$ and $W$ with sample size 200
%  at the significance level $\alpha=0.05$.
%           }\label{figure2}
%\end{figure}

%\begin{figure}
%  \centering
%  % Requires \usepackage{graphicx}
%\includegraphics[width=\textwidth]{introduction2.eps}\\
%  \caption{ The empirical power curves of Fan and Li's (1996) test and
%  Delgado and  Manteiga's (2001) test against the dimensions of $X$ and $W$ with sample size 200 at the significance level $\alpha=0.10$.
%           }\label{figure2}
%\end{figure}

%\begin{figure}
%  \centering
%  % Requires \usepackage{graphicx}
%\includegraphics[width=\textwidth]{Power.eps}\\
%  \caption{ The empirical power curves of Fan and Li's (1996) test and
%  Delgado and  Manteiga's (2001) test against the dimensions of $X$ and $W$ with sample size 200 at the significance level $\alpha=0.10$.
%           }\label{figure2}
%\end{figure}

\begin{table}[htb!]\caption{Empirical sizes and powers of $T^{DEE}_{n}$, $T^{FL}_{n}$ and $T^{DM}_{n}$
 with $p_1=2$ and $p_2=2$ in Example~1. \label{table1}
\vspace{0.75cm}}
\centering
 {\scriptsize\hspace{12.5cm}
\renewcommand{\arraystretch}{1}\tabcolsep 0.25cm
\begin{tabular}{ccccccccccc}
\hline
&&\multicolumn{3}{c}{$T^{DEE}_{n}$}
& \multicolumn{3}{c}{$T^{FL}_{n}$}
& \multicolumn{3}{c}{$T^{DM}_{n}$}\\
\hline
&$a/n$ &50 &100& 200 &50 &100& 200 &50 &100& 200\\
\hline
$X \sim N(0,\Sigma_1)$
&0  &0.0515&0.0485&0.0515&0.0570&0.0595&0.0550&0.0490&0.0520&0.0545\\
&0.4&0.1060&0.1535&0.4530&0.1560&0.2465&0.4560&0.3665&0.8490&0.9710\\
$W \sim N(0,\Sigma_1)$
&0.8&0.4060&0.7320&0.9230&0.3075&0.5880&0.9335&0.4420&0.8935&0.9860\\
&1.2&0.6020&0.8770&0.9570&0.4550&0.7750&0.9655&0.4485&0.9075&0.9805\\
&1.6&0.6680&0.9045&0.9695&0.5115&0.8700&0.9780&0.4590&0.9050&0.9890\\
&2.0&0.7320&0.9195&0.9785&0.5415&0.8945&0.9890&0.4375&0.9135&0.9935\\
 \hline
$X \sim N(0,\Sigma_2)$
&0  &0.0400&0.0635&0.0515&0.0615&0.0790&0.0580&0.0475&0.0470&0.0460\\
&0.4&0.0935&0.1450&0.3600&0.1610&0.2345&0.4230&0.3670&0.8520&0.9770\\
$W \sim N(0,\Sigma_2)$
&0.8&0.3435&0.6770&0.9015&0.3780&0.5430&0.9295&0.4400&0.9055&0.9880\\
&1.2&0.5655&0.8365&0.9420&0.5265&0.7085&0.9510&0.4410&0.9005&0.9930\\
&1.6&0.6650&0.8985&0.9520&0.6020&0.8340&0.9795&0.4480&0.9015&0.9890\\
&2.0&0.7110&0.9015&0.9675&0.6145&0.8980&0.9910&0.4645&0.9035&0.9960\\
 \hline
$X \sim N(0,\Sigma_3)$
&0  &0.0515&0.0525&0.0520&0.0590&0.0640&0.0555&0.0455&0.0535&0.0465\\
&0.4&0.0985&0.1155&0.3825&0.1225&0.2440&0.3270&0.3660&0.8495&0.9750\\
$W \sim N(0,\Sigma_3)$
&0.8&0.3640&0.7455&0.9380&0.3035&0.5905&0.8015&0.4275&0.8935&0.9870\\
&1.2&0.5930&0.8675&0.9470&0.4150&0.8190&0.9355&0.4455&0.9095&0.9865\\
&1.6&0.6830&0.9070&0.9640&0.5185&0.8685&0.9660&0.4605&0.8950&0.9875\\
&2.0&0.7495&0.9175&0.9700&0.5495&0.9055&0.9765&0.4925&0.9045&0.9880\\
 \hline
\end{tabular}
}
\end{table}

\begin{table}[htb!]\caption{Empirical sizes and powers of $T^{DEE}_{n}$, $T^{FL}_{n}$ and $T^{DM}_{n}$
 with $p_1=4$ and $p_2=4$ in Example~1. \label{table2}
\vspace{0.75cm}}
\centering
 {\scriptsize\hspace{12.5cm}
\renewcommand{\arraystretch}{1}\tabcolsep 0.25cm
\begin{tabular}{ccccccccccc}
\hline
&&\multicolumn{3}{c}{$T^{DEE}_{n}$}
& \multicolumn{3}{c}{$T^{FL}_{n}$}
& \multicolumn{3}{c}{$T^{DM}_{n}$}\\
\hline
&$a/n$ &50 &100& 200 &50 &100& 200 &50 &100& 200\\
\hline
$X \sim N(0,\Sigma_1)$
&0  &0.0545&0.0435&0.0520&0.0310&0.0620&0.0670&0.0020&0.0045&0.0240\\
&0.4&0.0730&0.1690&0.3770&0.0520&0.0615&0.0805&0.0060&0.0325&0.2205\\
$W \sim N(0,\Sigma_1)$
&0.8&0.3420&0.7355&0.9190&0.0550&0.0695&0.0955&0.0055&0.0480&0.2645\\
&1.2&0.5635&0.8660&0.9465&0.0545&0.0700&0.0950&0.0090&0.0425&0.2485\\
&1.6&0.6290&0.8860&0.9610&0.0530&0.0790&0.0960&0.0065&0.0520&0.2545\\
&2.0&0.6960&0.9115&0.9825&0.0550&0.0870&0.1150&0.0025&0.0485&0.2610\\
 \hline
$X \sim N(0,\Sigma_2)$
&0  &0.0620&0.0550&0.0525&0.0330&0.0600&0.0565&0.0250&0.0440&0.0800\\
&0.4&0.0815&0.1550&0.3320&0.0685&0.0820&0.0930&0.1260&0.2005&0.5100\\
$W \sim N(0,\Sigma_2)$
&0.8&0.3555&0.6950&0.9165&0.0810&0.1005&0.1205&0.1410&0.3930&0.7040\\
&1.2&0.5390&0.8295&0.9540&0.0900&0.1065&0.1415&0.1270&0.4260&0.7420\\
&1.6&0.6330&0.8795&0.9615&0.1000&0.1215&0.1430&0.1400&0.4285&0.7600\\
&2.0&0.6965&0.9090&0.9795&0.1070&0.1360&0.1595&0.1410&0.4180&0.7720\\
 \hline
$X \sim N(0,\Sigma_3)$
&0  &0.0400&0.0420&0.0555&0.0320&0.0705&0.06255&0.0070&0.0280&0.0715\\
&0.4&0.0680&0.1325&0.3900&0.0660&0.0720&0.07500&0.0410&0.1265&0.6300\\
$W \sim N(0,\Sigma_3)$
&0.8&0.3325&0.6960&0.9170&0.0815&0.0775&0.09505&0.0520&0.2320&0.6640\\
&1.2&0.5500&0.8345&0.9490&0.0830&0.0915&0.10005&0.0530&0.2315&0.6545\\
&1.6&0.6245&0.8845&0.9535&0.0865&0.1040&0.11850&0.0630&0.2320&0.6520\\
&2.0&0.6740&0.8890&0.9775&0.0875&0.1030&0.11655&0.0560&0.2500&0.6430\\
 \hline
\end{tabular}
}
\end{table}

\begin{table}[htb!]\caption{Empirical sizes and powers of $T^{DEE}_{n}$, $T^{FL}_{n}$ and $T^{DM}_{n}$
 with $p_1=4$ and $p_2=4$ in Example~2. \label{table3}
\vspace{0.75cm}}
\centering
 {\scriptsize\hspace{12.5cm}
\renewcommand{\arraystretch}{1}\tabcolsep 0.25cm
\begin{tabular}{ccccccccccc}
\hline
&&\multicolumn{3}{c}{$T^{DEE}_{n}$}
& \multicolumn{3}{c}{$T^{FL}_{n}$}
& \multicolumn{3}{c}{$T^{DM}_{n}$}\\
\hline
&$a/n$ &50 &100& 200 &50 &100& 200 &50 &100& 200\\
\hline
$X \sim N(0,\Sigma_1)$
&0  &0.0575&0.0520&0.0495&0.0360&0.0320&0.0695&0.0060&0.0100&0.0180\\
&0.4&0.0805&0.2225&0.6120&0.0665&0.0780&0.0910&0.0080&0.0520&0.1225\\
$W \sim N(0,\Sigma_1)$
&0.8&0.2910&0.7200&0.9190&0.0775&0.0875&0.0915&0.0040&0.0560&0.1790\\
&1.2&0.5465&0.8295&0.9405&0.0885&0.0820&0.1075&0.0200&0.0655&0.2290\\
&1.6&0.6215&0.8445&0.9685&0.0775&0.0915&0.1110&0.0160&0.0805&0.2415\\
&2.0&0.6565&0.8695&0.9730&0.0830&0.0960&0.1220&0.0260&0.0940&0.2615\\
 \hline
$X \sim N(0,\Sigma_2)$
&0  &0.0435&0.0435&0.0470&0.0640&0.0695&0.0760&0.0240&0.0380&0.0470\\
&0.4&0.1145&0.3035&0.6545&0.0745&0.0860&0.1005&0.0620&0.1255&0.5870\\
$W \sim N(0,\Sigma_2)$
&0.8&0.3735&0.6780&0.8755&0.0820&0.1065&0.1230&0.0845&0.2380&0.7190\\
&1.2&0.5570&0.7850&0.8830&0.0780&0.1000&0.1420&0.1070&0.2565&0.7660\\
&1.6&0.6445&0.8130&0.9070&0.0905&0.1090&0.1545&0.1150&0.3070&0.7610\\
&2.0&0.6745&0.8455&0.9300&0.0890&0.1265&0.1620&0.1115&0.3330&0.7900\\
 \hline
$X \sim N(0,\Sigma_3)$
&0  &0.0475&0.0490&0.0525&0.0360&0.0820&0.0705&0.0130&0.0230&0.0405\\
&0.4&0.0970&0.2465&0.6405&0.0590&0.0845&0.0920&0.0370&0.1140&0.4645\\
$W \sim N(0,\Sigma_3)$
&0.8&0.3360&0.7120&0.9120&0.0615&0.0850&0.1045&0.0580&0.1835&0.5715\\
&1.2&0.5600&0.7995&0.9265&0.0520&0.0925&0.1180&0.0690&0.2000&0.6370\\
&1.6&0.6625&0.8405&0.9370&0.0680&0.0990&0.1260&0.0700&0.2020&0.6125\\
&2.0&0.6760&0.8605&0.9500&0.0715&0.0950&0.1215&0.0730&0.2225&0.6240\\
 \hline
\end{tabular}
}
\end{table}

\begin{table}[htb!]\caption{Empirical sizes and powers of $T^{DEE}_{n}$, $T^{FL}_{n}$ and $T^{DM}_{n}$
 with $p_1=6$ and $p_2=6$ in Example~2. \label{table4}
\vspace{0.75cm}}
\centering
 {\scriptsize\hspace{12.5cm}
\renewcommand{\arraystretch}{1}\tabcolsep 0.25cm
\begin{tabular}{ccccccccccc}
\hline
&&\multicolumn{3}{c}{$T^{DEE}_{n}$}
& \multicolumn{3}{c}{$T^{FL}_{n}$}
& \multicolumn{3}{c}{$T^{DM}_{n}$}\\
\hline
&$a/n$ &50 &100& 200 &50 &100& 200 &50 &100& 200\\
\hline
$X \sim N(0,\Sigma_1)$
&0  &0.0405&0.0455&0.0465&0.0005&0.0005&0.0055&0&0&0\\
&0.4&0.0720&0.1860&0.5515&0.0005&0.0025&0.0035&0&0&0\\
$W \sim N(0,\Sigma_1)$
&0.8&0.2445&0.6240&0.9045&     0&0.0010&0.0030&0&0&0\\
&1.2&0.4575&0.7925&0.9190&     0&0.0015&0.0035&0&0&0\\
&1.6&0.5595&0.8260&0.9240&     0&0.0005&0.0035&0&0&0\\
&2.0&0.6255&0.8455&0.9445&     0&0.0015&0.0040&0&0&0\\
 \hline
$X \sim N(0,\Sigma_2)$
&0  &0.0425&0.0470&0.0485&0.0035&0.0135&0.0215&0.0020&0.0110&0.0195\\
&0.4&0.1335&0.2490&0.5550&0.0020&0.0080&0.0270&0.0020&0.0250&0.1040\\
$W \sim N(0,\Sigma_2)$
&0.8&0.3160&0.6215&0.8120&0.0020&0.0185&0.0275&0.0030&0.0270&0.1550\\
&1.2&0.5025&0.7645&0.8580&0.0030&0.0100&0.0250&0.0030&0.0290&0.1695\\
&1.6&0.5770&0.7950&0.8850&0.0050&0.0105&0.0290&0.0020&0.0280&0.1725\\
&2.0&0.6375&0.8290&0.9105&0.0070&0.0140&0.0205&0.0020&0.0320&0.1685\\
 \hline
$X \sim N(0,\Sigma_3)$
&0  &0.0455&0.0460&0.0475&     0&0.0025&0.0100&     0&0.0035&0.0115\\
&0.4&0.0985&0.2430&0.5830&0.0005&0.0035&0.0090&     0&0.0040&0.0520\\
$W \sim N(0,\Sigma_3)$
&0.8&0.3225&0.6515&0.8750&0.0030&0.0020&0.0095&0.0010&0.0075&0.0605\\
&1.2&0.4915&0.7695&0.8940&0.0015&0.0015&0.0080&0.0040&0.0080&0.0780\\
&1.6&0.5880&0.8030&0.9135&0.0010&0.0005&0.0060&0.0020&0.0120&0.0720\\
&2.0&0.6265&0.8380&0.9395&0.0010&0.0035&0.0085&0.0050&0.0135&0.0895\\
 \hline
\end{tabular}
}
\end{table}

\begin{table}[htb!]\caption{Empirical sizes and powers of $T^{DEE}_{n}$, $T^{FL}_{n}$ and $T^{DM}_{n}$
 with $p_1=4$ and $p_2=4$ in Example~3. \label{table5}
\vspace{0.75cm}}
\centering
 {\scriptsize\hspace{12.5cm}
\renewcommand{\arraystretch}{1}\tabcolsep 0.25cm
\begin{tabular}{ccccccccccc}
\hline
&&\multicolumn{3}{c}{$T^{DEE}_{n}$}
& \multicolumn{3}{c}{$T^{FL}_{n}$}
& \multicolumn{3}{c}{$T^{DM}_{n}$}\\
\hline
&$a/n$ &50 &100& 200 &50 &100& 200 &50 &100& 200\\
\hline
$X \sim N(0,\Sigma_1)$
&0  &0.0420&0.0530&0.0525&0.0575&0.0770&0.0760&0.0015&0.0080&0.0190\\
&0.4&0.0655&0.0985&0.3755&0.0510&0.0740&0.0765&0.0015&0.0280&0.1495\\
$W \sim N(0,\Sigma_1)$
&0.8&0.1470&0.5635&0.8930&0.0600&0.0845&0.1070&0.0040&0.0350&0.2085\\
&1.2&0.3690&0.8045&0.9390&0.0510&0.0840&0.0980&0.0035&0.0420&0.2525\\
&1.6&0.5730&0.8765&0.9485&0.0555&0.0840&0.1080&0.0090&0.0330&0.2510\\
&2.0&0.6480&0.8960&0.9645&0.0565&0.1000&0.1225&0.0030&0.0280&0.2595\\
 \hline
$X \sim N(0,\Sigma_2)$
&0  &0.0510&0.0545&0.0510&0.0720&0.0745&0.0810&0.0180&0.0440&0.0580\\
&0.4&0.0755&0.1165&0.2820&0.0685&0.0915&0.0945&0.0715&0.2460&0.3295\\
$W \sim N(0,\Sigma_2)$
&0.8&0.2075&0.5265&0.8490&0.0800&0.1175&0.1315&0.0925&0.3525&0.7430\\
&1.2&0.4245&0.7610&0.9365&0.0860&0.1040&0.1575&0.0950&0.3740&0.7615\\
&1.6&0.5595&0.8600&0.9480&0.0810&0.1095&0.1470&0.1145&0.3885&0.7630\\
&2.0&0.6315&0.8730&0.9560&0.0885&0.1205&0.1610&0.1140&0.3910&0.8185\\
 \hline
$X \sim N(0,\Sigma_3)$
&0  &0.0513&0.0550&0.0495&0.0515&0.0405&0.0790&0.0110& 0.0220&0.0340\\
&0.4&0.0750&0.1255&0.3135&0.0625&0.0720&0.0915&0.0200& 0.1320&0.4925\\
$W \sim N(0,\Sigma_3)$
&0.8&0.1945&0.5490&0.8840&0.0625&0.0875&0.1060&0.0340& 0.1870&0.6035\\
&1.2&0.4330&0.7795&0.9300&0.0655&0.0945&0.1270&0.0230& 0.1990&0.6100\\
&1.6&0.5955&0.8370&0.9460&0.0580&0.0960&0.1225&0.0470& 0.1960&0.6035\\
&2.0&0.6405&0.8610&0.9535&0.0595&0.0990&0.1290&0.0460& 0.2200&0.6280\\
 \hline
\end{tabular}
}
\end{table}

\begin{table}[htb!]\caption{Empirical sizes and powers of $T^{DEE}_{n}$ and the frequency of structure dimension $\hat{q}$ with $p_1=4$ and $p_2=4$  in Example~4. \label{table7}
\vspace{0.75cm}}
\centering
 {\footnotesize\hspace{12.5cm}
\renewcommand{\arraystretch}{1}\tabcolsep 0.35cm
\begin{tabular}{llllllll}
\hline
\multicolumn{1}{c}{$n$}
&\multicolumn{1}{c}{$a$}
&\multicolumn{1}{c}{$T^{DEE}_{n}$}
&\multicolumn{1}{c}{$T^{FL}_{n}$}
&\multicolumn{1}{c}{$T^{DM}_{n}$}\\%
%&\multicolumn{1}{c}{$\hat{q}>2$}\\
 \hline
 $n=200$
&0  &0.0555&0.0600&0.0290\\%&0.2320\\
&1  &0.3315&0.1455&0.1870\\%&0.6875\\
\hline
 $n=400$
&0&0.0535&0.0645&0.0465\\%&0.0870\\
&1&0.4910&0.1710&0.3205\\%&0.7915\\
 \hline
\end{tabular}
}
\end{table}

%\begin{table}[htb!]\caption{Result of $T^{DEE}_{n}$ for Significance Testing for Baseball hitters' salary data. \label{table8}
%\vspace{0.5cm}}
%\centering
% {\footnotesize\hspace{12.5cm}
%\renewcommand{\arraystretch}{1}\tabcolsep 0.5cm
%\begin{tabular}{ccccc}
%\hline
%&\multicolumn{2}{c}{Case I}
%& \multicolumn{2}{c}{Case II}\\
%\hline
%Bandwidth & $T^{DEE}_{n}$ & $P-$values & $T^{DEE}_{n}$ & $P-$values\\
%1.00&5.4597&0.0000&1.2814 &0.1000\\
%1.25&5.8197&0.0000&1.2171 &0.1118\\
%1.50&6.1806&0.0000&1.1517 &0.1247\\
%1.75&6.5831&0.0000&1.1171 &0.1320\\
%2.00&7.0327&0.0000&1.0968 &0.1364\\
%2.25&7.4862&0.0000&1.0775 &0.1406\\
%2.50&7.8843&0.0000&1.0429 &0.1485\\
%\hline
%\end{tabular}
%}
%\end{table}
%

\end{document}